\documentclass[sigconf,nonacm,screen]{acmart}

\usepackage{graphicx}
\usepackage{microtype}
\usepackage{epstopdf}
\usepackage{flushend}
\usepackage{balance}
\usepackage{xcolor}
\usepackage{verbatim}
\usepackage{listings}
\usepackage{booktabs}
\usepackage{nicefrac}
\usepackage{paralist}
\usepackage{subcaption}
\usepackage{algpseudocode}
\usepackage{soul}
\usepackage{subfiles}
\usepackage{multirow}
\usepackage{enumitem}
\usepackage{tabularx}
\usepackage{xspace}
\newcommand{\corobase}{CoroBase\xspace}

\def\thepaperkeywords{Transactions, cooperative multitasking, coroutine, memory stall}
\def\thepapertitle{\corobase: Coroutine-Oriented Main-Memory Database Engine}

\usepackage{float}
\hypersetup{
  pageanchor=true,
  plainpages=false,
  pdfpagelabels,
  bookmarks,
  bookmarksnumbered,
  pdfborder=0 0 0,  
  colorlinks=true,
  linkcolor=ACMRed,
  anchorcolor=ACMRed,
  citecolor=Blue, 
  pdftitle={\thepapertitle},
  pdfauthor={Yongjun He, Jiacheng Lu and Tianzheng Wang},
  pdfsubject={},
  pdfkeywords={\thepaperkeywords},
}

\usepackage{tikz}
\newcommand*\circled[1]{\tikz[baseline=(char.base)]{
            \node[shape=circle,fill=.,inner sep=0pt] (char) {\color{-.}\textsf\footnotesize #1};}}

\usepackage{breakurl}

\usepackage{algorithm}
\newfloat{algorithm}{t}{lop}

\definecolor{BrickRed}{rgb}{0.8, 0.25, 0.33}
\definecolor{Blue}{rgb}{0.01, 0.28, 1.0}

\floatstyle{ruled}
\definecolor{comment-red}{rgb}{1,0,0}

\newcommand\vldbdoi{10.14778/3430915.3430932}
\newcommand\vldbpages{XXX-XXX}
\newcommand\vldbvolume{14}
\newcommand\vldbissue{3}
\newcommand\vldbyear{2021}
\newcommand\vldbauthors{\authors}
\newcommand\vldbtitle{\shorttitle} 
\newcommand\vldbavailabilityurl{https://github.com/sfu-dis/corobase/tree/v1.0}
\newcommand\vldbpagestyle{empty} 

\newcommand{\rvalue}{\texttt{rvalue}\xspace}
\newcommand{\resume}{\texttt{resume}\xspace}
\newcommand{\prefetch}{\texttt{prefetch}\xspace}
\newcommand{\suspend}{\texttt{suspend}\xspace}
\newcommand{\coawait}{\texttt{co\_await}\xspace}
\newcommand{\coreturn}{\texttt{co\_return}\xspace}
\newcommand{\multiget}{\texttt{multi\_get}\xspace}
\newcommand{\get}{\texttt{get}\xspace}

\newcommand{\amac}{\texttt{AMAC-MK}\xspace}
\newcommand{\naive}{\texttt{Na\"{i}ve}\xspace}
\newcommand{\corofn}{\texttt{CORO-FN-MK}\xspace}
\newcommand{\ermia}{\texttt{ERMIA}\xspace}
\newcommand{\coro}{\texttt{CORO-MK}\xspace}
\newcommand{\cb}{\texttt{\corobase}\xspace}
\newcommand{\cbfn}{\texttt{\corobase-FN}\xspace}
\newcommand{\hybrid}{\texttt{Hybrid}\xspace}

\begin{document}

\title{\thepapertitle}
\subtitle{To appear in VLDB 2021}

\lstset {
language=C,
basicstyle=\ttfamily\small,
keywordstyle=\ttfamily\bfseries\color{blue},
commentstyle=\color{OliveGreen},
numbers=left,
numberstyle=\small,
numbersep=5pt,
tabsize=1,
gobble=0,
stepnumber=2,
xleftmargin=15pt,
escapeinside={(@*}{*@)},
morekeywords={},
columns=fullflexible,
}

\author{Yongjun He}
\affiliation{
  \institution{Simon Fraser University}
}
\email{yongjunh@sfu.ca}

\author{Jiacheng Lu}
\affiliation{
  \institution{Simon Fraser University}
}
\email{jiacheng_lu@sfu.ca}

\author{Tianzheng Wang}
\affiliation{
  \institution{Simon Fraser University}
}
\email{tzwang@sfu.ca}

\begin{abstract}
Data stalls are a major overhead in main-memory database engines due to the use of pointer-rich data structures. 
Lightweight coroutines ease the implementation of software prefetching to hide data stalls by overlapping computation and asynchronous data prefetching. 
Prior solutions, however, mainly focused on (1) individual components and operations and (2) intra-transaction batching that requires interface changes, breaking backward compatibility.
It was not clear how they apply to a full database engine and how much end-to-end benefit they bring under various workloads.

This paper presents \corobase, a main-memory database engine that tackles these challenges with a new \textit{coroutine-to-transaction} paradigm.
Coroutine-to-transaction models transactions as coroutines and thus enables inter-transaction batching, avoiding application changes but retaining the benefits of prefetching.
We show that on a 48-core server, \corobase can perform close to 2$\times$ better for read-intensive workloads and remain competitive for workloads that inherently do not benefit from software prefetching.
\end{abstract}

\maketitle

\pagestyle{\vldbpagestyle}
\begingroup\small\noindent\raggedright\textbf{PVLDB Reference Format:}\\
\vldbauthors. \vldbtitle. PVLDB, \vldbvolume(\vldbissue): \vldbpages, \vldbyear.\\
\href{https://doi.org/\vldbdoi}{doi:\vldbdoi}
\endgroup
\begingroup
\renewcommand\thefootnote{}\footnote{\noindent
This work is licensed under the Creative Commons BY-NC-ND 4.0 International License. Visit \url{https://creativecommons.org/licenses/by-nc-nd/4.0/} to view a copy of this license. For any use beyond those covered by this license, obtain permission by emailing \href{mailto:info@vldb.org}{info@vldb.org}. Copyright is held by the owner/author(s). Publication rights licensed to the VLDB Endowment. \\
\raggedright Proceedings of the VLDB Endowment, Vol. \vldbvolume, No. \vldbissue\ %
ISSN 2150-8097. \\
\href{https://doi.org/\vldbdoi}{doi:\vldbdoi} \\
}\addtocounter{footnote}{-1}\endgroup

\ifdefempty{\vldbavailabilityurl}{}{
\vspace{.3cm}
\begingroup\small\noindent\raggedright\textbf{PVLDB Artifact Availability:}\\
The source code, data, and/or other artifacts have been made available at \url{\vldbavailabilityurl}.
\endgroup
}

\section{Introduction}
\label{sec:intro}

\begin{figure}[t]
\centering
\includegraphics[width=\columnwidth]{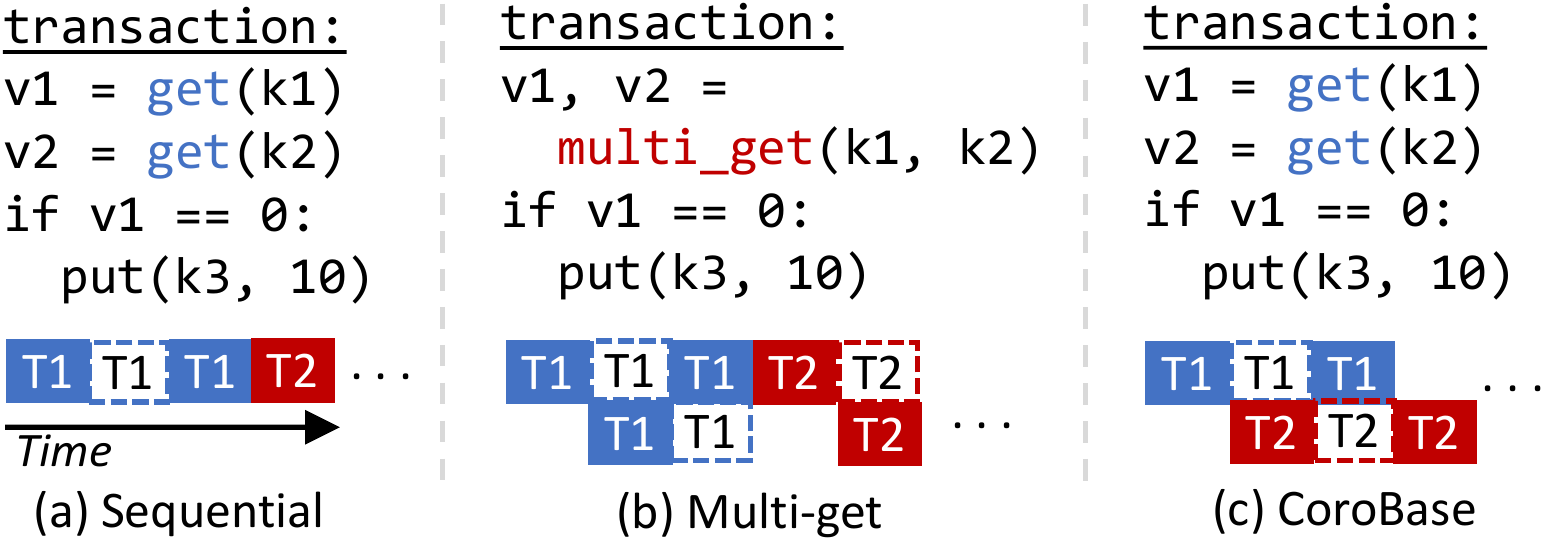}
\caption{\textmd{Data access interfaces and execution behavior under (a) sequential execution without interleaving, 
(b) prior approaches that require applications use multi-key interfaces, 
(c) \corobase which hides data stalls and maintains backward compatibility.}}
\label{fig:coro-vs-others}
\end{figure}

Modern main-memory database engines~\cite{Hyper,Hekaton,HStore,ERMIA,Silo,Deuteronomy,FOEDUS,Cicada} use memory-optimized data structures~\cite{ART,HOT,BwTree,Masstree} to offer high performance on multicore CPUs. 
Many such data structures rely on pointer chasing~\cite{CompilerBasedPrefetching} which can stall the CPU upon cache misses.
For example, in Figure~\ref{fig:coro-vs-others}(a), to execute two \texttt{SELECT} (\texttt{get}) queries, the engine may traverse a tree, and if a needed tree node is not cache-resident, dereferencing a pointer to it stalls the CPU (dotted box in the figure) to fetch the node from memory.
Computation (solid box) would not resume until data is in the cache. 
With the wide speed gap between CPU and memory,
memory accesses have become a major overhead~\cite{boncz1999database,manegold2009database}. 
The emergence of capacious but slower persistent memory~\cite{Intel3DXP} is further widening this gap.

Modern processors allow multiple outstanding cache misses and provide \prefetch instructions~\cite{IntelManual} for software to explicitly bring data from memory to CPU caches.
This gave rise to software prefetching techniques~\cite{ChenHashJoin,AMAC,KillerNanosecond,InterleaveJoins,InterleaveJoinsVLDBJ,CompilerPrefetching,CompilerBasedPrefetching} that \textit{hide} memory access latency by overlapping data fetching and computation, alleviating pointer chasing overhead.
Most of these techniques, however, require hand-crafting asynchronous/pipelined algorithms or state machines to be able to suspend/resume execution as needed. 
This is a difficult and error-prone process; the resulted code often deviates a lot from the original code, making it hard to maintain~\cite{KillerNanosecond}. 

\subsection{Software Prefetching via Coroutines} 
With the recent standardization in C++20~\cite{Coroutine}, coroutines greatly ease the implementation of software prefetching.
Coroutines~\cite{moura2009revisiting} are functions that can suspend voluntarily and be resumed later. 
Functions that involve pointer chasing can be written as coroutines which are executed (interleaved) in batches. 
Before dereferencing a pointer in coroutine $t1$, the thread issues a \prefetch followed by a \suspend to pause $t1$ and switches to another coroutine $t2$, overlapping data fetching in $t1$ and computation in $t2$.

Compared to earlier approaches~\cite{ChenHashJoin,AMAC}, coroutines only require \prefetch/\suspend be inserted into sequential code, greatly simplifying implementation while delivering high performance, as the switching overhead can be cheaper than a last-level cache miss~\cite{KillerNanosecond}.
However, adopting software prefetching remains challenging.

First, existing approaches typically use intra-transaction batching which mandates multi-key interfaces that can break backward compatibility. 
For example, in Figure~\ref{fig:coro-vs-others}(b) an application\footnote{The ``application'' may be another database system component or an end-user application that uses the record access interfaces provided by the database engine.} 
uses \multiget to retrieve a batch of records at once in a transaction.
Cache misses caused by probing \texttt{k1} (\texttt{k2}) in a tree are hidden behind the computation part of probing \texttt{k2} (\texttt{k1}).
While intra-transaction batching is a natural fit for some operators (e.g., IN-predicate queries~\cite{InterleaveJoinsVLDBJ,InterleaveJoins}), 
it is not always directly applicable.
Changing the application is not always feasible and may not achieve the desired improvement as depending requests need to be issued in separate batches, limiting interleaving opportunities.
Short (or even single-record) transactions also cannot benefit much due to the lack of interleaving opportunity.
It would be desirable to allow batching operations across transactions, i.e., inter-transaction batching. 

Second, prior work provided only piece-wise solutions, focusing on optimizing individual database operations (e.g., index traversal~\cite{KillerNanosecond} and hash join~\cite{ChenHashJoin,InterleaveJoins}).
Despite the significant improvement (e.g., up to 3$\times$ faster for tree probing~\cite{KillerNanosecond}), it was not clear how much overall improvement one can expect when these techniques are applied in a full database engine that involves various components.

Overall, these issues lead to two key questions: 
\begin{itemize}[leftmargin=*]\setlength\itemsep{0em}
\item How should a database engine adopt coroutine-based software prefetching, preferably without requiring application changes?
\item How much end-to-end benefit can software prefetching bring to a database engine under realistic workloads?
\end{itemize}

\subsection{\corobase}
To answer these questions, 
we propose and evaluate \corobase, a multi-version, main-memory database engine that uses coroutines to hide data stalls.
The crux of \corobase is a simple but effective \textit{coroutine-to-transaction} paradigm that models transactions as coroutines, to enable \textit{inter}-transaction batching and maintain backward compatibility. 
Worker threads receive transaction requests and switch among transactions (rather than requests within a transaction) without requiring intra-transaction batching or multi-key interfaces.
As Figure~\ref{fig:coro-vs-others}(c) shows, the application remains unchanged as batching and interleaving happen at the transaction level.

Coroutine-to-transaction can be easily adopted to hide data stalls in different database engine components and can even work together with multi-key based approaches.
In particular, in multi-version systems versions of data records are typically chained using linked lists~\cite{MVCCEval}, traversing which constitutes another main source of data stalls, in addition to index traversals.
\corobase transparently suspends and resumes transactions upon pointer dereferences during version chain traversals. 
This way, \corobase ``coroutinizes'' the full data access paths to provide an end-to-end solution.

To explore how coroutine-to-transaction impacts common design principles of main-memory database systems, instead of building \corobase from scratch, we base it on ERMIA~\cite{ERMIA}, an open-source, multi-version main-memory database engine. 
This allows us to devise an end-to-end solution and explore how easy (or hard) it is to adopt coroutine-to-transaction in an existing engine, which we expect to be a common starting point for most practitioners.
In this context, we discuss solutions to issues brought by coroutine-to-transaction, such as (nested) coroutine switching overhead, 
higher latency and more complex resource management in later sections. 


On a 48-core server, our evaluation results corroborate with prior work and show that software prefetching is mainly (unsurprisingly) beneficial to read-dominant workloads, with close to 2$\times$ improvement over highly-optimized baselines. 
For write-intensive workloads, we find mixed results with up to 45\% improvement depending on access patterns.
Importantly, \corobase retains competitive performance for workloads that inherently do not benefit from prefetching, thanks to its low-overhead coroutine design. 

Note that our goal is \textit{not} to outperform prior work, but to 
(1) effectively adopt software prefetching in a database engine without necessitating new interfaces, and 
(2) understand its end-to-end benefits.
Hand-crafted techniques usually present the performance upper bound; \corobase strikes a balance between performance, programmability and backward compatibility.

\subsection{Contributions and Paper Organization}
We make four contributions. 
\circled{1} We highlight the challenges for adopting software prefetching in main-memory database engines. 
\circled{2} We propose a new execution model, coroutine-to-transaction, to enable inter-transaction batching and avoid interface changes while retaining the benefits of prefetching.
\circled{3} We build \corobase, a main-memory multi-version database engine that uses coroutine-to-transaction to hide data stalls during index and version chain traversals.
We explore the design tradeoffs by describing our experience of transforming an existing engine to use coroutine-to-transaction. 
\circled{4} We conduct a comprehensive evaluation of \corobase to quantify the end-to-end effect of prefetching under various workloads. 
\corobase is open-source at \url{https://github.com/sfu-dis/corobase}.

Next, we give the necessary background in Section~\ref{sec:bg}. 
Sections~\ref{sec:principles}--~\ref{sec:cb} then present the design principles and details of \corobase.
Section~\ref{sec:eval} quantifies the end-to-end benefits of software prefetching.
We cover related work in Section~\ref{sec:related-work} and conclude in Section~\ref{sec:conclusion}.

\section{Background} 
\label{sec:bg}
This section gives the necessary background on software prefetching techniques and coroutines to motivate our work.

\subsection{Software Prefetching}
Although modern CPUs use sophisticated hardware prefetching mechanisms, they are not effective on reducing pointer-chasing overheads, due to the irregular access patterns in pointer-intensive data structures.
For instance, when traversing a tree, it is difficult for hardware to predict and prefetch correctly the node which is going to be accessed next, until the node is needed right away.

\textbf{Basic Idea.} Software prefetching techniques~\cite{AMAC,ChenHashJoin,InterleaveJoins,KillerNanosecond} use workload semantics to issue \prefetch instructions~\cite{IntelManual} to explicitly bring data into CPU caches.
Worker threads handle requests (e.g., tree search) in batches.
To access data (e.g., a tree node) in request $t1$ which may incur a cache miss, the thread issues a \prefetch and switches to another request $t2$, and repeats this process.
While the data needed by $t1$ is being fetched from memory to CPU cache, the worker thread handles $t2$, which may further cause the thread to issue \prefetch and switch to another request.
By the time the worker thread switches back to $t1$, the hope is that the needed data is (already and still) cache-resident.
The thread then picks up at where it left for $t1$, dereferences the pointer to the prefetched data and continues executing $t1$ until the next possible cache miss upon which a \prefetch will be issued.
It is important that the switching mechanism and representation of requests are cheap and lightweight enough to achieve a net gain.

\textbf{Hand-Crafted Approaches.} The mechanism we just described fits naturally with many loop-based operations. 
Group prefetching and software pipelined prefetching~\cite{ChenHashJoin} overlap multiple hash table lookups to accelerate hash joins.
After a \prefetch is issued, the control flow switches to execute the computation stage of another operation.
Asynchronous memory access chaining (AMAC)~\cite{AMAC} is a general approach that allows one to transform a heterogeneous set of operations into state machines to facilitate switching between operations upon cache misses.
A notable drawback of these approaches is they require developers hand-craft algorithms. 
The resulted code is typically not intuitive to understand and hard to maintain. 
This limits their application to simple or individual data structures and operations (e.g., tree traversal).
Recent approaches tackle this challenge using lightweight coroutines, described next. 

\subsection{Coroutines}
\label{subsec:coroutine}
Coroutines~\cite{conway1963design} are generalizations of functions with two special characteristics: 
(1) During the execution between invoke and return, a coroutine can suspend and be resumed at manually defined points. 
(2) Each coroutine preserves local variables until it is destroyed. 
Traditional stackful coroutines~\cite{moura2009revisiting} use separate runtime stacks to keep track of local variables and function calls.
They have been available as third-party libraries~\cite{Boost,Fibers} and are convenient to use, but exhibit high overhead that is greater than the cost of a cache miss~\cite{KillerNanosecond}, defeating the purpose of hiding memory stalls.

\begin{figure}[t]
\centering
\includegraphics[width=\columnwidth]{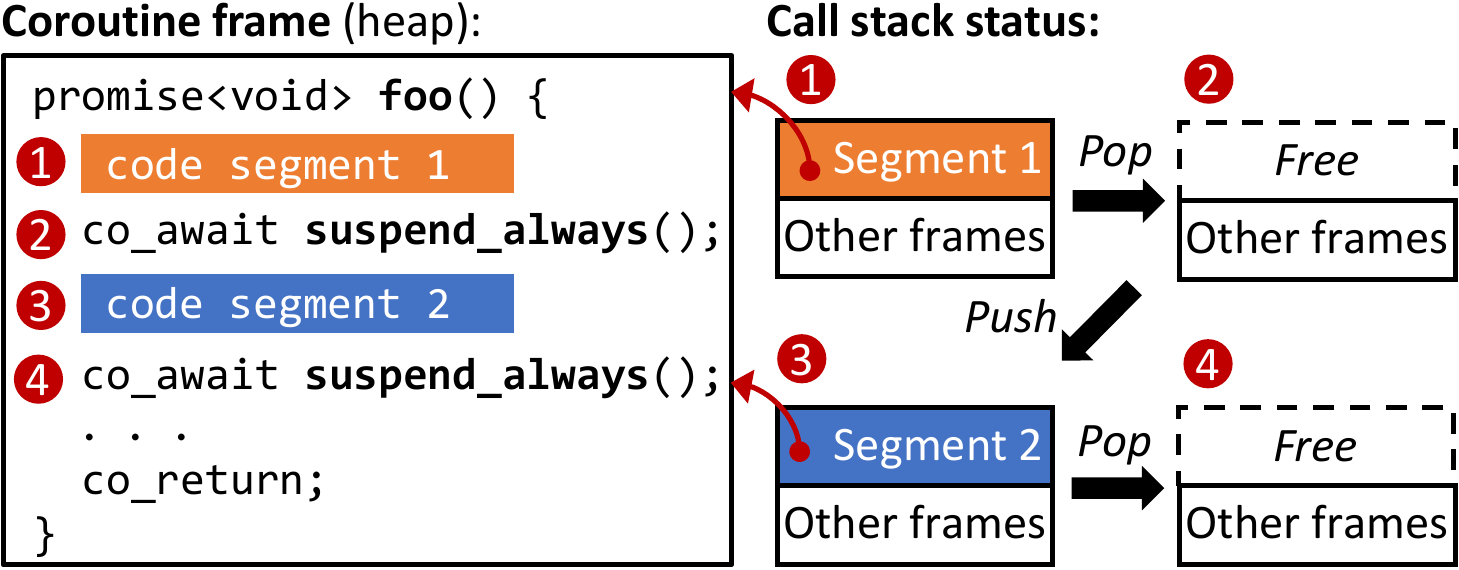}
\caption{\textmd{Stackless coroutine that directly uses the underlying thread's stack. Necessary states (e.g., local variables and return value) are maintained in dynamically allocated memory.}}
\label{fig:coroutine}
\end{figure}

\textbf{Stackless Coroutines.} Recent stackless coroutines standardized in C++20~\cite{Coroutine} (which is our focus) exhibit low overhead in construction and context switching\footnote{Not to be confused with context switches at the OS level. 
In main-memory systems, threads are typically pinned to mitigate the impact of OS scheduling. Throughout this paper ``contexts'' refers to coroutines/transactions that are pure user-space constructs.} (cheaper than a last-level cache miss).
They do not own stacks and run on the call stack of the underlying thread.
Invoking a coroutine is similar to invoking a normal function, but its states (e.g., local variables that live across suspension points) are kept in dynamically allocated memory (coroutine frames) that survive \suspend/\resume cycles.
Figure~\ref{fig:coroutine} shows an example in C++20: any function that uses coroutine keywords (e.g., \texttt{co\_await}, \texttt{co\_return}) is a coroutine.
A coroutine returns a \texttt{promise\_type} structure that allows querying the coroutine's states, such as whether it is completed and its return value.
The \texttt{co\_await} keyword operates on a \texttt{promise\_type} and is translated by the compiler into a code block that can save the states in a coroutine frame and pop the call stack frame.
The \texttt{suspend\_always} object is an instance of \texttt{promise\_type} that has no logic and suspends unconditionally.
The \texttt{co\_return} keyword matches the syntax of \texttt{return}, but instead of returning an \rvalue, it stores the returned value into a coroutine frame.
As Figure~\ref{fig:coroutine} shows, upon starting (step \circled{1}) or resuming (step \circled{3}) a coroutine, a frame is created and pushed onto the stack.
At unconditional suspension points (steps \circled{2} and \circled{4}), the frame is popped and control is returned to the caller.
Since the coroutine frame lives on the heap, coroutine states are still retained after the stack frame is popped.
Coroutine frames need to be explicitly destroyed after the coroutine finishes execution.


\textbf{Scheduling.}
Each worker thread essentially runs a scheduler that keeps switching between coroutines, such as the one below:
\begin{small}
\begin{verbatim}
1. // construct coroutines
2. for i = 0 to batch_size - 1:
3.   coroutine_promises[i] = foo(...);
4. // switch between active coroutines
5. while any(coroutine_promises, x: not x.done()):
6.   for i = 0 to batch_size - 1:
7.     if not coroutine_promises[i].done():
8.       coroutine_promises[i].resume()
\end{verbatim} 
\end{small}
After creating a batch of operations (coroutines) at lines 1--2, it invokes and switches among coroutines (lines 4--8).
The \texttt{batch\_size} parameter determines the number of inflight memory fetches and how effectively memory stalls can be hidden: once a coroutine suspends, it is not resumed before the other \texttt{batch\_size-1} coroutines are examined. 
Prior work has shown that the optimal \texttt{batch\_size} is roughly the number of outstanding memory accesses that can be supported by the CPU (10 in current Intel x86 processors)~\cite{InterleaveJoins}.

\textbf{Nested Stackless Coroutines.}
Similar to ``normal'' functions, a coroutine may invoke, suspend and resume another coroutine using \texttt{co\_await}, forming a chain of nested coroutines. 
By default, when a stackless coroutine suspends, control is returned to its caller.
Real-world systems often employ deep function calls for high-level operations to modularize their implementation. 
To support interleaving at the operation level in database engines (e.g., search), a mechanism that allows control to be returned to the top-level (i.e., the scheduler) is necessary. 
This is typically done by returning control level-by-level, from the lowest-level suspending coroutine to the scheduler, through every stack frame.
Note that the number of frames preceding the suspending coroutine on the stack may not be the same as its position in the call chain.
For the first \suspend in the coroutine chain, the entire chain is on the stack.
For subsequent {\suspend}s, however, the stack frames start from the last suspended coroutine instead of the top level one, since it is directly resumed by the scheduler.
When a coroutine $c$ finishes execution, the scheduler resumes execution of $c$'s parent coroutine.
As a result, a sequential program with nested function calls can be easily transformed into nested coroutine calls by adding \prefetch and \suspend statements. 

Astute reader may have noticed that this approach can be easily used to realize coroutine-to-transaction.
However, doing so can bring non-trivial overhead associated with scheduling and maintaining coroutine states; we discuss details in later sections. 



\subsection{Coroutine-based Software Prefetching in Main-Memory Database Engines}
\label{subsec:bg-coro}
Modern main-memory database systems eliminate I/O operations from the critical path. 
This allows worker threads to execute each transaction without any interruptions. 

\begin{figure}[t]
\centering
\includegraphics[width=\columnwidth]{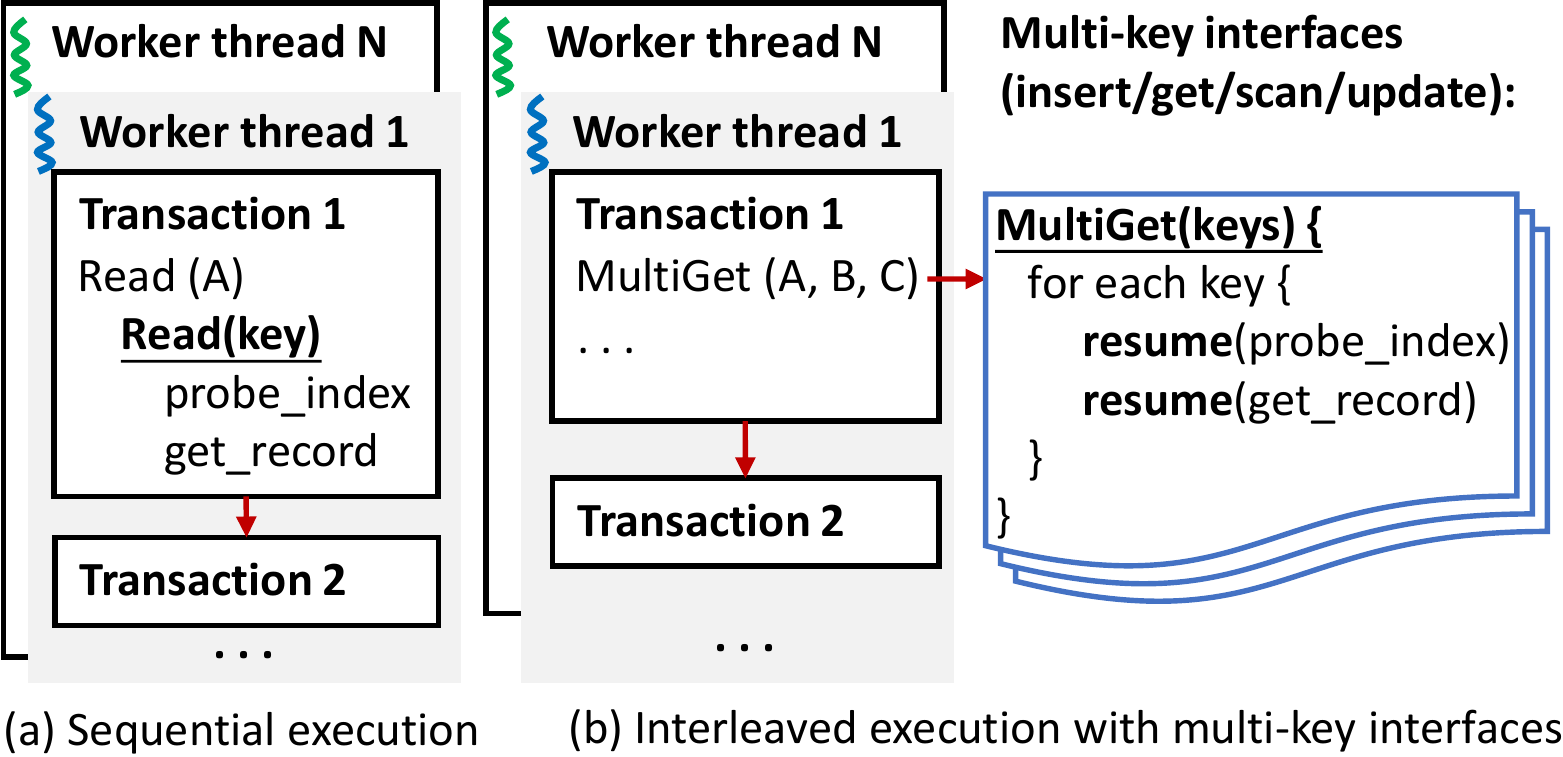}
\caption{\textmd{Sequential vs. interleaving under thread-to-transaction.}}
\label{fig:exec-models}
\end{figure}

\textbf{Execution Model.} 
Recent studies have shown that data stalls are a major overhead in both OLTP and OLAP workloads~\cite{OLAPAnalysis,sirin2016micro}.
In this paper, we mainly focus on OLTP workloads.
With I/O off the critical path, thread-to-transaction has been the dominating execution model in main-memory environments for transaction execution. 
Each worker thread executes transactions one by one without context switching to handle additional transactions unless the current transaction concluded (i.e., committed or aborted). 
Figure~\ref{fig:exec-models}(a) shows an example of worker threads executing transactions under this model. 
To read a record, the worker thread sequentially executes the corresponding functions that implement the needed functionality, including (1) probing an index to learn about the physical address of the target record, and (2) fetching the record from the address. 
After all the operations of the current transaction (\texttt{Transaction 1} in the figure) are finished, the worker thread continues to serve the next transaction.

\textbf{Software Prefetching under Thread-to-Transaction.} With nested coroutines, it is straightforward to transform individual operations to use software prefetching, by adding suspend points into existing sequential code.
However, under thread-to-transaction, once a thread starts to work on a transaction, it cannot switch to another. 
As a result, the caller of these operations now essentially runs a scheduler that switches between individual operations, i.e., using intra-transaction batching.
In the case of a transaction reading records, for example, in Figure~\ref{fig:exec-models}(b), the transaction calls a \multiget function that accepts a set of keys as its parameter and runs a scheduler that switches between coroutines that do the heavylifting of record access and may suspend upon cache misses.
All these actions happen in the context of a transaction; another transaction can only be started after the current transaction being handled concludes, limiting inter-transaction batching and necessitating interface changes that may break backward compatibility. 

\section{Design Principles}
\label{sec:principles}
We summarize four desirable properties and principles that should be followed when designing coroutine-based database engines:

\begin{itemize}[leftmargin=*]\setlength\itemsep{0em}
\item \textbf{Maintain Backward Compatibility.} The engine should allow applications to continue to use single-key interfaces. Interleaving should be enabled within the engine without user intervention.
\item \textbf{Low Context Switching Overhead.} It should be at least lower than the cost of a last-level cache miss to warrant any end-to-end performance improvement in most cases. For workloads that do not have enough data stalls to benefit from prefetching, having low switching overhead can help retain competitive performance.
\item \textbf{Maximize Batching Opportunities.} The batching mechanism should allow both intra- and inter-transaction interleaving. This would allow arbitrary query to benefit from prefetching, in addition to operators that naturally fits the batching paradigm.
\item \textbf{Easy Implementation.} A salient feature of coroutine is it only needs simple changes to sequential code base; a new design must retain this property for maintainability and programmability.
\end{itemize}

\section{\corobase Design}
\label{sec:cb}
Now we describe the design of \corobase, a multi-version main-memory database engine based on the coroutine-to-transaction execution model.
We do so by taking an existing memory-optimized database engine (ERMIA~\cite{ERMIA}) and transforming it to use coroutine-to-transaction.
As we mentioned in Section~\ref{sec:intro}, this allows us to contrast and highlight the feasibility and potential of coroutine-to-transaction, and reason about the programming effort required to adopt coroutine-to-transaction.
However, \corobase and coroutine-to-transaction can be applied to other systems. 


\begin{figure}[t]
\centering
\includegraphics[width=\columnwidth]{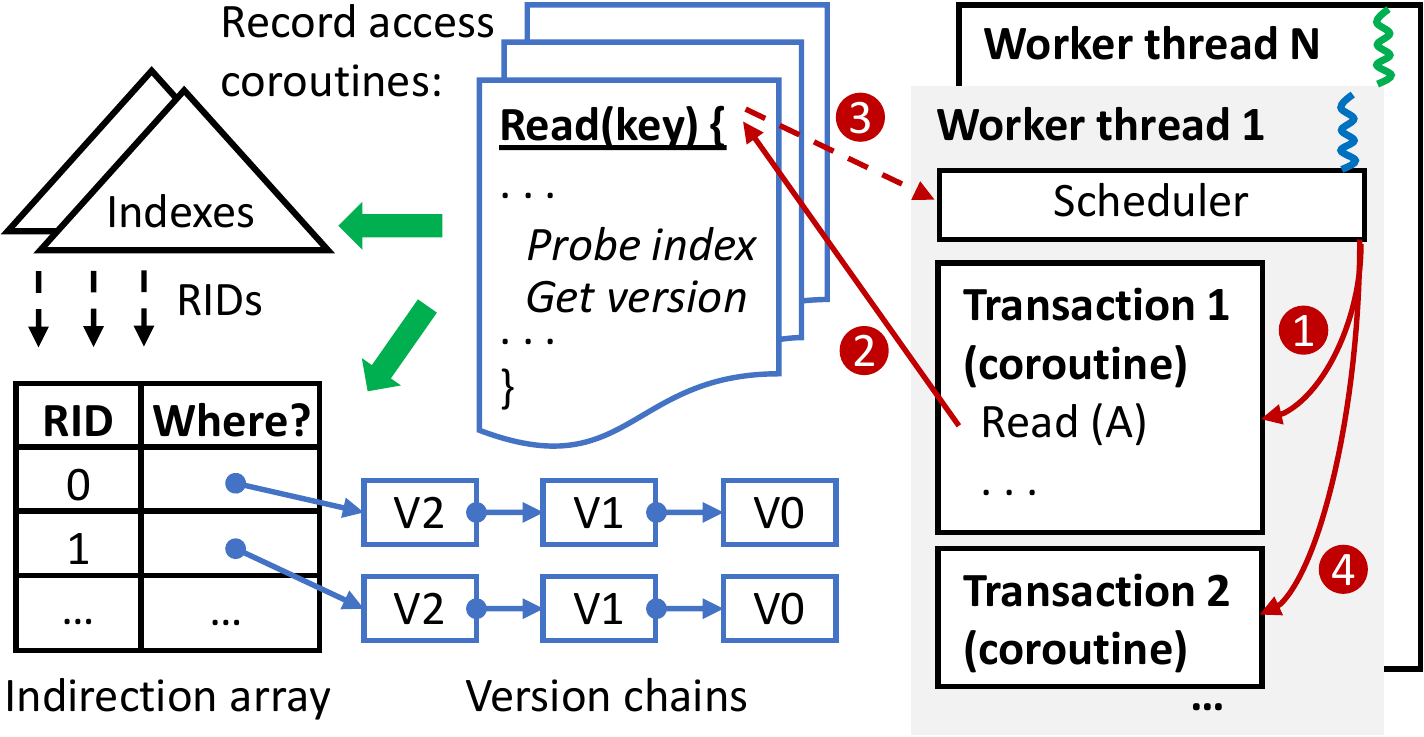}
\caption{\textmd{\corobase overview. 
Indexes map keys to record IDs (RIDs) which uniquely identify records. 
Record versions are maintained by version chains (linked lists).
Each worker thread runs a scheduler that \protect\circled{1} starts/resumes transactions (coroutines). 
\protect\circled{2} A transaction may invoke other coroutines that implement specific operations, which may suspend as needed but \protect\circled{3} return control directly to the scheduler, which can 
\protect\circled{4} resume a different transaction.}}
\label{fig:corobase}
\end{figure}

\subsection{Overview}
\corobase organizes data and controls data version visibility in ways similar to other main-memory multi-version systems~\cite{ERMIA,Cicada,Deuteronomy,Hekaton,MVCCEval} (in our specific case, ERMIA~\cite{ERMIA}).
Figure~\ref{fig:corobase} gives the overall design of \corobase.
For each record, \corobase maintains multiple versions that are chained together in a linked list, with the latest version (ordered by logical timestamps) as the list head.
This is a common design in multi-version systems~\cite{MVCCEval}.
Each record is uniquely identified by a logical record ID (RID) that never changes throughout the lifetime of the record, in contrast to physical RIDs in traditional disk-based systems~\cite{CowBook}. 
For each table, we maintain an indirection array~\cite{ERMIA,IndArraySSD} that maps RID to the virtual memory pointer to the record's latest version which further points to the next older version, and so on.
Indexes map keys to RIDs, instead of pointers to record versions. 
A main benefit of this approach is that record updates (i.e., creation of new versions) or movement (e.g., from memory to storage) will not always mandate secondary index updates. 
Same as ERMIA, \corobase uses Masstree~\cite{Masstree} for indexing and all data accesses are done through indexes, however, the choice of index types is orthogonal to the techniques being proposed here.

With the indirection and version chain design, accessing a record is a two-step process: the worker thread first traverses an index of the underlying table, and then consults the indirection array (using the RID found in the leaf node as index) to find the actual record and suitable version by traversing the version chain of that record.
We defer details on determining version visibility to later sections when we discuss concurrency control.
Under thread-to-transaction, this two-step process is done synchronously by the worker thread, with record access procedures implemented as multiple levels of functions.
Under coroutine-to-transaction, record access procedures are implemented as coroutines, instead of ``normal'' functions.
As Figure~\ref{fig:corobase} shows, each worker thread independently runs a scheduler that switches between a batch of transactions (coroutines).
The key to realize this model is transforming nested functions on the data access path into coroutines, which we elaborate next.


\subsection{Fully-Nested Coroutine-to-Transaction}
\label{sec:generic}
To support common data access operations (insert/read/update/scan/delete), a straightforward way is to transform function call chains that may cause cache misses into nested stackless coroutines outlined in Section~\ref{subsec:coroutine}.
Functions that will not incur cache misses may be kept as ``normal'' functions.
For index, we follow prior work~\cite{KillerNanosecond} to add \suspend statements (\texttt{suspend\_always} in C++20) as needed after each \prefetch, which can be identified easily as Masstree already prefetches nodes.\footnote{Based on \url{https://github.com/kohler/masstree-beta}.}
We use \coawait to invoke other coroutines and replace \texttt{return} with \coreturn.
For version chain traversal, we issue \prefetch and \suspend before dereferencing a linked list node pointer.
These changes are straightforward and only require inserting \prefetch/\suspend statements and replacing keywords.
Our implementation defines macros to automatically convert between function and coroutine versions of the code, easing code maintainability.\footnote{For example, the record read function/coroutine can be found at: \url{https://github.com/sfu-dis/corobase/blob/v1.0/ermia.cc\#L145}. The \texttt{AWAIT} and \texttt{PROMISE} macros transparently convert between coroutine and function versions.}
Thus, a call chain of N functions is replaced by an (up to) N-level coroutine chain. 
Coroutines at any level may voluntarily suspend, after which control goes back to the scheduler which resumes the next transaction that is not done yet; control then goes to the newly resumed transaction.
\begin{algorithm}[t]
\begin{lstlisting}[language=python,
	mathescape,
	gobble=0,
	keywordstyle=\ttfamily\bfseries\color{blue},
	]
def scheduler(batch_size):
  while not shutdown:
    [T] = get_transaction_requests()
    enter_epoch()
    while done < batch_size:
      done = 0
      for i = 0 to batch_size - 1:
        if T[i].is_done:
          ++done
        else
          T[i].resume()
    exit_epoch()
\end{lstlisting}

\caption{Scheduler for coroutine-to-transaction.}
\label{alg:scheduler}
\end{algorithm}

Figure~\ref{fig:corobase} shows how control flows end-to-end.
When a new transaction is created, \corobase starts to execute it in a coroutine (step \circled{1} in the figure).
Subsequent operations (e.g., read/write/scan) are further handled in the context of their corresponding transaction coroutine (step \circled{2}).
All the operations are also coroutines that may suspend and get resumed. 
For example, in Figure~\ref{fig:corobase}, the \texttt{Read} coroutine may further use other coroutines to traverse a tree index structure to find the requested record's RID, followed by invoking yet another coroutine that traverses the version chain to retrieve the desirable version.
Upon a possible cache miss, the executing coroutine (e.g., index traversal as part of a \texttt{Read} call) issues a \prefetch to bring the needed memory to CPU caches asynchronously, followed by a \suspend which returns control directly to the scheduler (step \circled{3}).
This allows the scheduler to further resume another transaction (step \circled{4}), hoping to overlap computation and data fetching.
After a transaction commits or aborts, its coroutine structures are destroyed and control is returned to the scheduler which may resume another active transaction. 
Finally, after every transaction in the batch is concluded, the scheduler starts a new batch of transactions.

Coroutine-to-transaction moves the responsibility of batching from the user API level to the engine level, by grouping transactions. 
Each worker thread runs a coroutine scheduler which accepts and handles transaction requests. 
In \corobase we use a round-robin scheduler shown in Algorithm~\ref{alg:scheduler}.
The scheduler function keeps batching and switching between incoming transactions (lines 7--11).
It loops over each batch to execute transactions. 
When a query in a transaction suspends, control returns to the scheduler which then resumes the next in-progress transaction (line 11).
Note that each time the scheduler takes a fixed number (denoted as \texttt{batch\_size}) of transactions, and when a transaction finishes, we do not start a new one until the whole batch is processed.
The rationale behind is to preserve locality and avoid overheads associated with initializing transaction contexts.
Although it may reduce the possible window of overlapping, we observe the impact is negligible.
Avoiding irregular, ad hoc transaction context initialization helps maintain competitive performance for workloads that inherently do not benefit from prefetching where the scheduler activities and switching are pure overheads that should be minimized.
Processing transactions in strict batches also eases the adoption of epoch-based resource management in coroutine environments, as Section~\ref{subsec:epoch} describes.
The downside is that individual transaction latency may become higher.
Our focus is OLTP where transactions are often similar and short, so we anticipate the impact to be modest. 
For workloads that may mix short transactions and long queries in a batch, other approaches, e.g., a scheduler that takes transaction priority into account when choosing the next transaction to resume may be more attractive for reducing system response time.

While easy to implement, fully-nested coroutine-to-transaction and coroutine-to-transaction in general bring three main challenges, which we elaborate next.

\subsection{Two-Level Coroutine-to-Transaction}
\label{subsec:2level}
Since currently there is no way for software to tell whether dereferencing a pointer would cause a cache miss~\cite{InterleaveJoins,KillerNanosecond}, software has to ``guess'' which memory accesses may cause a cache miss. 
\corobase issues \prefetch and \suspend upon dereferencing pointers to index nodes and record objects in version chains based on profiling results.
To reduce the cost of wrong guesses, it is crucial to reduce switching overheads. 
Database engine code typically uses nested, deep call chains of multiple functions to modularize implementation.
As Section~\ref{sec:eval} shows, blindly transforming deep function call chains into coroutine chains using fully-nested coroutine-to-transaction incurs non-trivial overhead that overshadows the benefits brought by prefetching.
Yet completely flattening (inlining) the entire call chain to become a single coroutine mixes application logic and database engine code, defeating the purpose of coroutine-to-transaction.

\corobase takes a middle ground to flatten only nested calls within the database engine, forming a two-level structure that balances performance and programmability.
This allows the application to still use the conventional interfaces; under the hood, transaction coroutines may invoke other coroutines for individual operations (e.g., \get), which are single-level coroutines with all the nested coroutines that may suspend inlined.
Sequential functions that do not suspend are not inlined unless the compiler does so transparently.
At a first glance, it may seem tedious or infeasible to inline the whole read/write paths in a database engine manually. 
However, this is largely mechanical and straightforward.
For instance, it took us shorter than three hours to flatten the search operation in Masstree~\cite{Masstree}, the index structure used by \corobase.
The flattened code occupies $<$100 lines and still largely maintains the original logic.\footnote{Details in our code repo at lines 375--469 at \url{https://github.com/sfu-dis/corobase/blob/v1.0/corobase.cc\#L375}.}
This shows that flattening functions is in fact feasible.
Moreover, there is a rich body of work in compilers about function inlining and flattening~\cite{EliminateStackSaveRAM,necula2002cil,rivoira2011compiler,IncrementalInlineJIT,OptTransformCatalogue} that can help automate this process.
For example, developers can still write small, modularized functions, but a source-to-source transformation pass can be performed to flatten the code before compilation.

The downside of flattening is that it may cause more instruction cache misses because the same code segment (previously short functions) may appear repeatedly in different coroutines.
For example, the same tree traversal code is required by both update and read operations.
Individual coroutines may become larger, causing more instruction cache misses.
However, as Section~\ref{sec:eval} shows, the benefits outweigh this drawback. 
Code reordering~\cite{AggressiveInline} can also be used as an optimization in compilers to reduce the instruction fetch overhead. 
Discussion of code transformation techniques is beyond the scope of this paper; we leave it as promising future work.

\subsection{Resource Management}
\label{subsec:epoch}
Resource management in the physical layer is tightly coupled with transaction execution model.
Under thread-to-transaction, ``transaction'' is almost a synonymy of thread, allowing transparent application of parallel programming techniques, in particular epoch-based memory reclamation and thread-local storage to improve performance. 
Most of these techniques are implicitly thread-centric, sequential algorithms that do not consider the possibility of coroutine switching.
Although the OS scheduler may deschedule a thread working on a task $t$, the thread does not switch to another task.
When the thread resumes, it picks up from where it left to continue executing $t$.
This implicit assumption brings extra challenges for coroutine-based asynchronous programming, which we describe and tackle next.
Note that these issues are not unique to database systems, and our solutions are generally applicable to other systems employing coroutines and parallel programming techniques.

\textbf{Epoch-based Reclamation.} 
Many main-memory database engines rely on lock-free data structures, e.g., lock-free lists~\cite{HarrisLockFreeLinkedList} and trees~\cite{BwTree}.
Threads may access memory that is simultaneously being removed from the data structure.
Although no new accesses are possible once the memory block is unlinked, existing accesses must be allowed to finish before the memory can be recycled. 

Epoch-based memory reclamation~\cite{ReclaimPerf} is a popular approach to implementing this. 
The basic idea is for each thread to register (enter an epoch) upon accessing memory, and ensure the unlinked memory block is not recycled until all threads in the epoch have deregistered (exited).
The epoch is advanced periodically depending pre-defined conditions, e.g., when the amount of allocated memory passes a threshold.
An assumption is that data access are coordinated by thread boundaries. 
Under thread-to-transaction, a transaction exclusively uses all the resources associated with a thread, so thread boundaries are also transaction boundaries.
Transactions can transparently use the epoch enter/exit machinery.
However, under coroutine-to-transaction this may lead to memory corruption:  
Suppose transactions $T1$ and $T2$ run on the same thread, and $T1$ has entered epoch $e$ before it suspends.
Now the scheduler switches to $T2$ which is already in epoch $e$ and issued epoch exit, allowing the memory to be freed, although it is still needed by $T1$ later. 

\corobase solves this problem by decoupling epoch enter/exit from transactions for the scheduler to issue them.
Upon starting/finishing the processing of a batch of transactions, the worker thread enters/exits the epoch (lines 4 and 12 in Algorithm~\ref{alg:scheduler}).
This fits nicely with our scheduling policy which only handles whole batches.
It also reduces the overhead associated with epoch-based reclamation as each thread registers/deregisters itself much less frequently.
Another potential approach is to implement nested enter/exit machinery that allows a thread to register multiple times as needed. 
Though flexible, this approach is error-prone to implement, and brings much higher bookkeeping overhead.

\textbf{Thread-Local Storage (TLS).}
TLS is widely used to reduce initialization and allocation overheads.
In particular, ERMIA uses thread-local read/write sets and log buffers, as well as thread-local scratch areas for storing temporaries such as records that were read and new data to be added to the database.
Logically, these structures are transaction-local.
Although making these structures TLS greatly reduces memory allocation overheads, it conflicts with the coroutine-to-transaction paradigm, which must decouple threads and transactions.
In other words, they need to be transaction-local to provide proper isolation among transactions.
To reduce the performance impact, we expand each individual TLS variable to become an array of variables, one per transaction in the batch, and store them again as TLS variables.
Upon system initialization, each worker thread creates the TLS array before starting to handle requests.
When a transaction starts (e.g., the $i$-th in a batch), it takes the corresponding TLS array entry for use. 
This way, we avoid allocation/initialization overhead similar to how it was done under thread-to-transaction, but provide proper isolation among transactions. 
The tradeoff is that we consume (\texttt{batch\_size} times) more memory space.
As we show in Section~\ref{sec:eval}, our approach makes a practical tradeoff as the optimal batch size does not exceed ten.

\subsection{Concurrency Control and Synchronization}
\label{subsec:sync-cc}

\corobase inherits the shared-everything architecture, synchronization and concurrency control protocols from ERMIA (snapshot isolation with the serial safety net~\cite{SSN} for serializability). 
A worker thread is free to access any part of the database.
Version visibility is controlled using timestamps drawn from a global, monotonically increasing counter maintained by the engine. 
Upon transaction start, the worker thread reads the timestamp counter to obtain a begin timestamp $b$.
When reading a record, the version with the latest timestamp that is smaller than $b$ is visible.
To update a record, the transaction must be able to see the latest version of the record.
To commit, the worker thread atomically increments the timestamp counter (e.g., using atomic \texttt{fetch-and-add}~\cite{IntelManual}) to obtain a commit timestamp which is written on the record versions created by it.

We observe that adopting coroutine-to-transaction in shared-everything systems required no change for snapshot isolation to work.
Similar to ERMIA, \corobase adopts the serial safety net (SSN)~\cite{SSN}, a certifier that can be applied on top of snapshot isolation to achieve serializability.
It tracks dependencies among transactions and aborts transactions that may lead to non-serializable execution.
Adapting SSN to \corobase mainly requires turning TLS bitmaps used for tracking readers in tuple headers~\cite{SSN} into transaction-local. 
This adds \texttt{batch\_size} bits per thread (compared to one in ERMIA). 
The impact is minimal because of the small ($\le$10) batch size, and stalls caused by bitmap accesses can be easily hidden by prefetching.
Devising a potentially more efficient approach under very high core count (e.g., 1000) is interesting future work.
Since \corobase allows multiple open transactions per thread, the additional overhead on supporting serializability may widen the conflict window and increase abort rate.
Our experiments show that the impact is very small, again because the desirable batch sizes are not big (4--8).



For physical-level data structures such as indexes and version chains, coroutines bring extra challenges if they use latches for synchronization (e.g., higher chance to deadlock with multiple transactions open on a thread).
However, in main-memory database engines these data structures mainly use optimistic concurrency without much (if not none of) locking. 
In \corobase and ERMIA, index (Masstree~\cite{Masstree}) traversals proceed without acquiring any latches and rely on versioning for correctness.
This makes it straightforward to coroutinize the index structure for read/scan and part of update/insert/delete operations (they share the same traversal code to reach the leaf level).
Locks are only held when a tree node is being updated.
Hand-over-hand locking is used during structural modification operations (SMOs) such as splits. 
Our profiling results show that cache misses on code paths that use hand-over-hand locking make up less than 6\% of overall misses under an insert-only workload.
It is not high enough to benefit much from prefetching, and will increase code complexity.
Atomic instructions such as compare-and-swap used by most latch implementations also do not benefit much from prefetching. 
Therefore, we do not issue \suspend on SMO code paths that use hand-over-hand locking.


More general, coroutine-centric synchronization primitives such as asynchronous mutex\footnote{Such as the \texttt{async\_mutex} in CppCoro: 
\url{https://github.com/lewissbaker/cppcoro/blob/1140628b6e9e6048234d404cc393d855ae80d3e7/include/cppcoro/async_mutex.hpp}.} are also being devised. 
How these primitives would apply to database systems remains to be explored.

\begin{table}[t]
\caption{\textmd{Necessary changes to adopt coroutine-to-transaction in systems based on thread-to-transaction. The key is to ensure proper isolation for transactions concurrently open on the same thread.}}
\label{tbl:changes}
\begin{tabular}{@{}p{1.85cm}p{6.15cm}@{}}
\toprule
\textbf{Component} &
  \textbf{Modifications} \\ \midrule
Concurrency Control (CC) &
  \begin{tabular}[c]{@{}p{6.15cm}@{}}{\it Shared-everything:} transparent, but need careful deadlock handling if pessimistic locking used.\\ {\it Shared-nothing:} re-introduce CC.\end{tabular} \\ \midrule
{Synchronization} &
  \begin{tabular}[c]{@{}l@{}}1. Thread-local to transaction-local.\\ 2. Avoid holding latches upon suspension.\end{tabular} \\ \midrule
Resource & 1. Thread-local to transaction-local.\\ 
Management &2. Piggyback on batching to reduce overhead.\\ \midrule
{Durability} &
  Transparent, with transaction-local log buffers. \\ \bottomrule
\end{tabular}%
\end{table}

\subsection{Discussions}
\label{subsec:discuss}
Coroutine-to-transaction only dictates how queries and transactions are interleaved.
It does not require fundamental changes to components in ERMIA. 
Table~\ref{tbl:changes} summarizes the necessary changes in ERMIA and engines that may make different assumptions than ERMIA's.
Beyond concurrency control, synchronization and resource management, we find that the durability mechanism (logging, recovery and checkpointing) is mostly orthogonal to the execution model. 
The only change is to transform the thread-local log buffer to become transaction-local, which is straightforward. 

Coroutine-to-transaction fits naturally with shared-everything, multi-versioned systems that use optimistic flavored concurrency control protocols.
For pessimistic locking, coroutine-to-transaction may increase deadlock rates if transactions suspend while holding a lock. 
Optimistic and multi-version approaches alleviate this issue as reads and writes do not block each other, although adding serializability in general widens the conflict window. 

Different from shared-everything systems, shared-nothing systems~\cite{HStore} partition data and restrict threads to only access its own data partition. 
This allowed vastly simplified synchronization and concurrency control protocols: in most cases if a transaction only accesses data in one partition, no synchronization or concurrency control is needed at all, as there is at most one active transaction at any time working on a partition. 
To adopt coroutine-to-transaction in shared-nothing systems, concurrency control and synchronization needs to be (re-)introduced to provide proper isolation between transactions running on the same thread.


Some systems~\cite{ReactDB,PWV,Bohm,DORA} explore intra-transaction parallelism to improve performance: a transaction is decomposed into pieces, each of which is executed by a thread dedicated to a partition of data, allowing non-conflicting data accesses in the same transaction to proceed in parallel.
Data stalls may still occur as threads use pointer-intensive data structures (e.g., indexes) to access data.
Coroutine-to-transaction can be adapted to model the individual pieces as coroutines to hide stalls. 
This would require changes such as a scheduler described in Algorithm~\ref{alg:scheduler} in the transaction executors in Bohm~\cite{Bohm} and ReactDB~\cite{ReactDB}); we leave these for future work.

Finally, \corobase removes the need for multi-key operations, but still supports them.
A transaction can call a multi-key operation which interleaves operations within a transaction and does not return control to the scheduler until completion.
A transaction can also use operations in both interfaces to combine inter- and intra-transaction interleaving. 
For example, it can invoke a \get, followed by an AMAC-based join to reduce latency and coroutine switching overhead.
This hybrid approach can be attractive when the efforts for changing interfaces is not high. 
Section~\ref{sec:eval} quantifies the potential of this approach.

\section{Evaluation}
\label{sec:eval}
Now we evaluate \corobase to understand the end-to-end effect of software prefetching under various workloads. 
Through experiments, we confirm the following:
\begin{itemize}[leftmargin=*]\setlength\itemsep{0em}
\item \corobase enables inter-transaction batching to effectively batch arbitrary queries to benefit from software prefetching. 
\item In addition to read-dominant workloads, \corobase also improves on read-write workloads while remaining competitive for workloads that inherently do not benefit from software prefetching.
\item \corobase can improve performance with and without hyperthreading, on top of hardware prefetching.
\end{itemize}

\subsection{Experimental Setup}
We use a dual-socket server equipped with two 24-core Intel Xeon Gold 6252 CPUs clocked at 2.1GHz (up to 3.7GHz with turbo boost).
The CPU has 35.75M last-level cache. 
In total the server has 48 cores (96 hyperthreads) and 384GB main memory occupying all the six channels per socket to maximize memory bandwidth.
We compile all the code using Clang 10 with coroutine support on Arch Linux with Linux kernel 5.6.5.
All the data is kept in memory using \texttt{tmpfs}. 
We report the average throughput and latency numbers of three 30-second runs of each experiment.

\textbf{System Model.} 
Similar to prior work~\cite{Silo,FOEDUS,ERMIA}, we implement benchmarks in C++ directly using APIs exposed by the database engine, without SQL or networking layers. 
Using coroutines to alleviate overheads in these layers is promising but orthogonal work.
The database engine is compiled as a shared library, which is then linked with the benchmark code to perform tests.

\textbf{Variants.}
We conduct experiments using the following variants which are all implemented based on ERMIA~\cite{ERMIA}.\footnote{ERMIA code downloaded from \url{https://github.com/sfu-dis/ermia}.} 

\begin{itemize}[leftmargin=*]\setlength\itemsep{0em}
\item \naive: Baseline that uses thread-to-transaction and executes transactions sequentially without interleaving or prefetching.
\item \ermia: Same as \naive but with \prefetch instructions carefully added to index and version chain traversal code.
\item \amac: Same as \ermia but applications use hand-crafted multi-key interfaces based on AMAC.
\item \coro: Same as \ermia but applications use multi-key interfaces based on flattened coroutines.
\item \corofn: Same as \coro but with fully-nested coroutines described in Section~\ref{subsec:coroutine}.
\item \cbfn: \corobase that uses the fully-nested coroutine-to-transaction design. No changes in applications.
\item \cb: Same as \cbfn but uses the optimized 2-level coroutine-to-transaction design described in Section~\ref{subsec:2level}.
\item \hybrid: Same as \cb but selectively leverages multi-key interfaces in TPC-C transactions (details in Section~\ref{subsec:tpcc}).
\end{itemize}

We use a customized allocator to avoid coroutine frame allocation/deallocation bottlenecks.
Hardware prefetching is enabled for all runs.
For interleaved executions, we experimented with different batch sizes and use the optimal setting (eight) unless specified otherwise. 
We use snapshot isolation\footnote{
\label{fn:ssn}
We also ran experiments under the serializable isolation level (using SSN on top of snapshot isolation). 
The results show that SSN adds a fixed amount of overhead ($\sim$10--15\%, similar to the numbers reported earlier for thread-to-transaction systems~\cite{SSN}). 
We therefore focus on experiments under SI for clarity.}
for all runs except for pure index-probing workloads which do not involve transactions (described later).
We also perform experiments with and without hyperthreading to explore its impact.

\subsection{Benchmarks}
We use both microbenchmarks and standard benchmarks to stress test and understand the end-to-end potential of \corobase.

\textbf{Microbenchmarks.} We use YCSB~\cite{YCSB} to compare in detail the impact of different design decisions.
The workload models point read, read-modify-write (RMW), update and scan transactions on a single table with specified access patterns.
We use a $\sim$15GB database of one billion records with 8-byte keys and 8-byte values.

\textbf{Standard Benchmarks.}
We use TPC-C~\cite{TPCC} to quantify the end-to-end benefits of software prefetching under \corobase.
To show comprehensively how \corobase performs under a realistic and varying set of workloads with different read/write ratios, we run both the original TPC-C and two variants, TPC-CR~\cite{QueryFresh} and TPC-CH~\cite{ERMIA}.
TPC-CR is a simplified read-only version of TPC-C that comprises 50\% of StockLevel and 50\% OrderStatus transactions. 
TPC-CH adds a modified version of the Query2 transaction (Q2*) in TPC-H~\cite{TPCH} to TPC-C's transaction mix.
This makes TPC-CH a heterogeneous workload that resembles hybrid transaction-analytical processing (HTAP) scenarios.\footnote{
\label{fn:tpcch}
Our focus is on OLTP (Section~\ref{subsec:bg-coro}).
We run TPC-CH to explore how \corobase performs under various read-intensive workloads.
Hiding data stalls in OLAP workloads requires further investigation that considers various access patterns and constraints.} 
We use the same implementation in ERMIA~\cite{ERMIA} where the transaction picks a random region and updates records in the stock table whose quantity is lower than a pre-defined threshold. 
The size of Q2* is determined mainly by the portion of the suppliers table it needs to access. 
We modify the transaction mix to be 10\% of Q2*, 40\% of NewOrder, 38\% of Payment, plus 4\% of StockLevel, Delivery and OrderStatus each.
For all TPC-C benchmarks, we set scale factor to 1000 (warehouses).
Each transaction uniform-randomly chooses and works on their home warehouse, but 1\% of New-Order and 15\% of Payment transactions respectively access remote warehouses.

\subsection{Sequential vs. Interleaved Execution}
\label{subsec:probe}
\begin{figure}[t]
\centering
\includegraphics[width=\columnwidth]{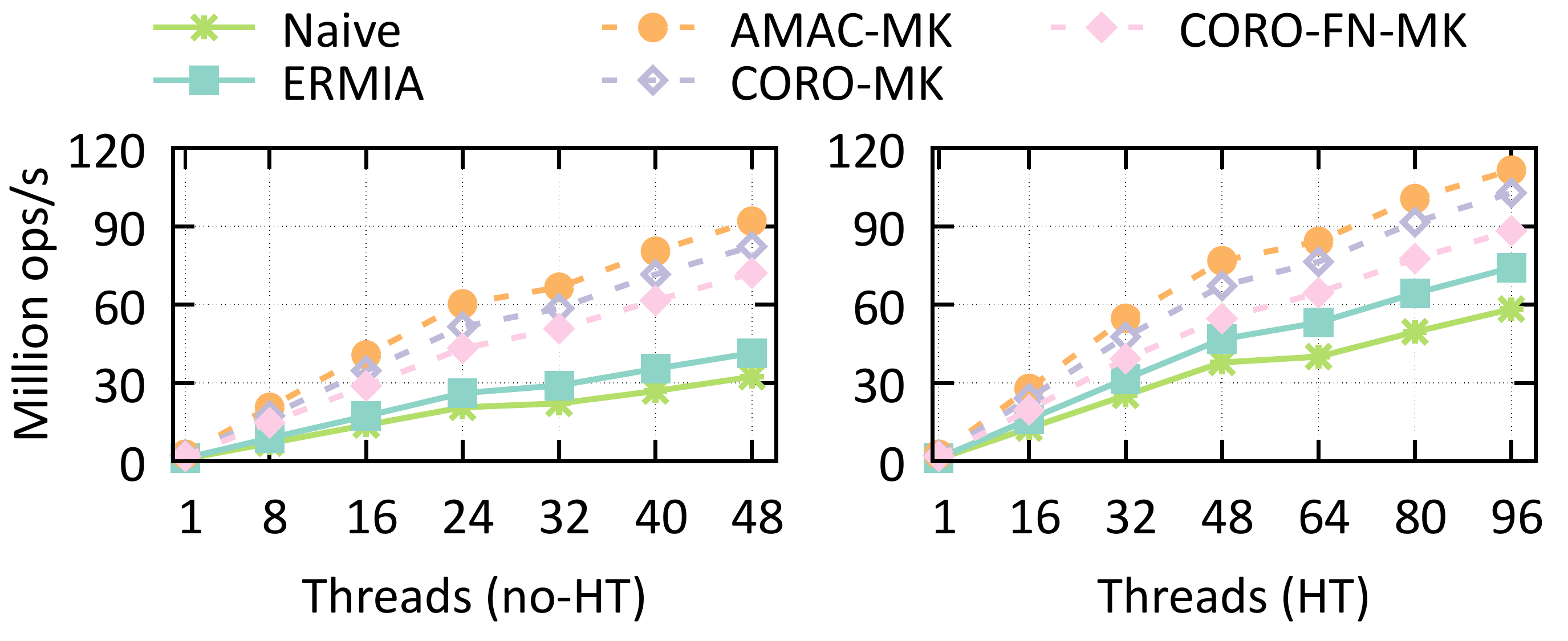}
\caption{\textmd{Throughput of index probing using multi-get with hyperthreading disabled (left) and enabled (right). 
Fully-nested coroutines (\corofn) are $\sim$30\% slower than AMAC. 
Flattened coroutines (\coro) reduce the gap to 8--10\% under high concurrency.}}
\label{fig:probe-only}
\end{figure}

As mentioned in earlier sections, our goal in this paper is to adopt coroutine-based interleaving in a database engine and understand its end-to-end benefits. 
Therefore, it is important to set the proper expectation on the possible gain that could be achieved by \corobase. 
To do this, we run a simple non-transactional index probing workload where worker threads keep issuing \multiget requests.
Each \multiget issues 10 requests against the index but does not access the actual database record. 
The workload is similar to what prior work~\cite{KillerNanosecond} used to evaluate index performance.
As Figure~\ref{fig:probe-only} shows, on average \amac outperforms \naive/\ermia by up to 2.3$\times$/2.96$\times$ without hyperthreading. 
Since \amac does not incur much overhead in managing additional metadata like coroutine frames, these results set the upper bound of the potential gain of interleave execution.
However, \amac uses highly-optimized but complex, hand-crafted code, making it much less practical.
Using single-level coroutines, \coro achieves up to 2.56$\times$/1.99$\times$ higher throughput than \naive/\ermia, which is 17\% faster than \corofn because of its lower switching overhead.
With hyperthreading, the improvement becomes smaller across all variants. 

These results match what was reported earlier in the literature, and set the upper bound for \cb to be up to $\sim$2$\times$ faster than optimized sequential execution  that already uses prefetching (i.e., the \ermia variant), or $\sim$2.5$\times$ faster than \naive under read-only workloads.
In the rest of this section, we mark the upper bound and multi-key variants that require interface changes as dashed lines in figures, and explore how closely \corobase matches the upper bound under various workloads.

\subsection{Effect of Coroutine-to-Transaction}
\label{subsec:coro-tx-effect}
\begin{figure}[t]
\centering
\includegraphics[width=\columnwidth]{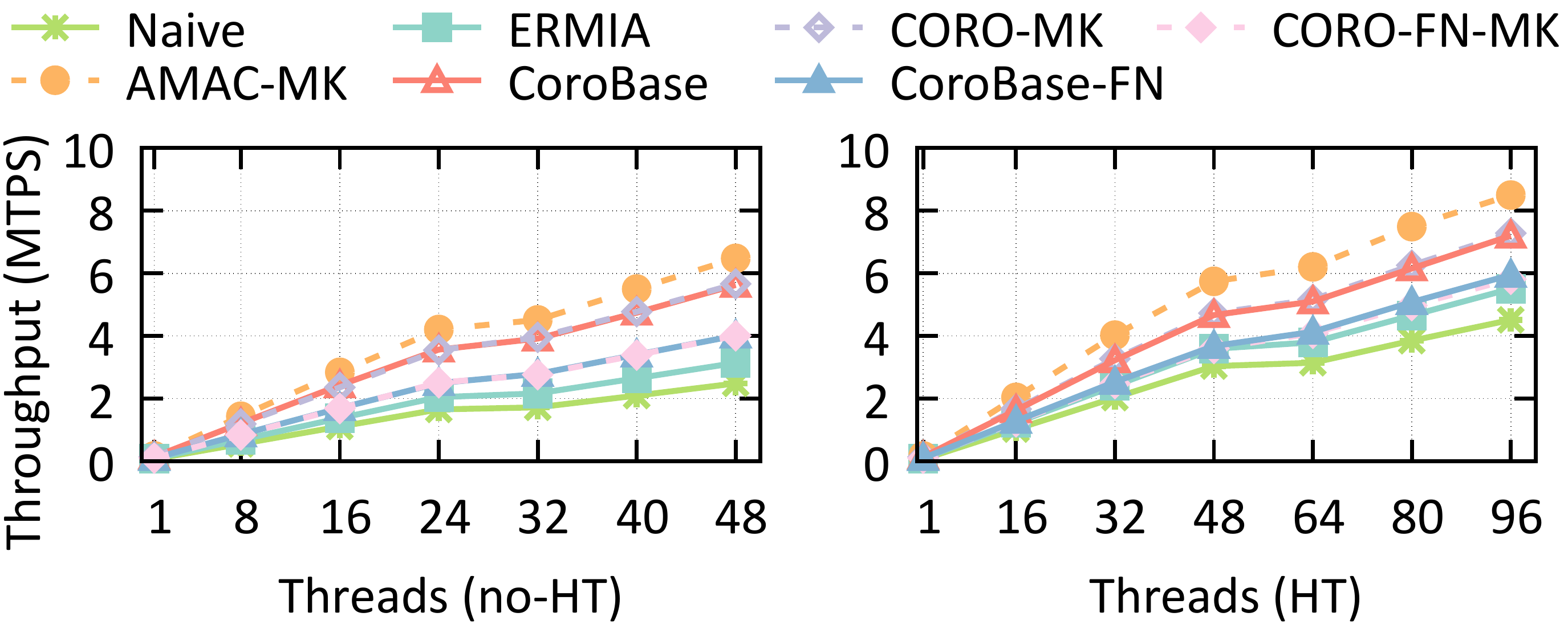}
\caption{\textmd{Throughput of a read-only YCSB workload (10 reads per transaction) without (left) and with (right) hyperthreading. \corobase matches the performance of multi-key variants but without requiring application changes.}}
\label{fig:read-only}
\end{figure}

Our first end-to-end experiment evaluates the effectiveness of coroutine-to-transaction.
We use a read-only YCSB workload where each transaction issues 10 record read operations that are uniform randomly chosen from the database table; we expand on to other operations (write and scan) later.
Note that different from the previous probe-only experiment in Section~\ref{subsec:probe}, from now on we run fully transactional experiments that both probe indexes and access database records.
We compare variants that use coroutine-to-transaction (\cb and \cbfn) with other variants that use highly-optimized multi-key interfaces and thread-to-transaction.
Figure~\ref{fig:read-only} plots result using physical cores (left) and hyperthreads (right).
Compared to the baselines (\ermia and \naive), all variants show significant improvement. 
Because of the use of highly-optimized, hand-crafted state machines and multi-key interfaces, \amac outperforms all the other variants.
Without hyperthreading, \cb exhibits an average of $\sim$15/$\sim$18\% slowdown from \amac, but is still $\sim$1.3/1.8$\times$ faster than \ermia when hyperthreading is enabled/disabled.
This is mainly caused by the inherent overhead of the coroutine machinery.
\corofn and \cbfn are only up $\sim$1.25$\times$ faster than \ermia due to high switching overhead.
\coro and \cb minimize switching overhead by flattening the entire record access call chain (index probing and version chain traversal).
The only difference is \coro uses \multiget, while \cb allows the application to remain unchanged using single-key interfaces.
Therefore, it is necessary to flatten call chains as much as possible to reduce context switching overhead.
As the coroutine infrastructure continues to improve, we expect the gap between \cb, \cbfn and \amac to become smaller.

\begin{figure}[t]
\centering
\includegraphics[width=\columnwidth]{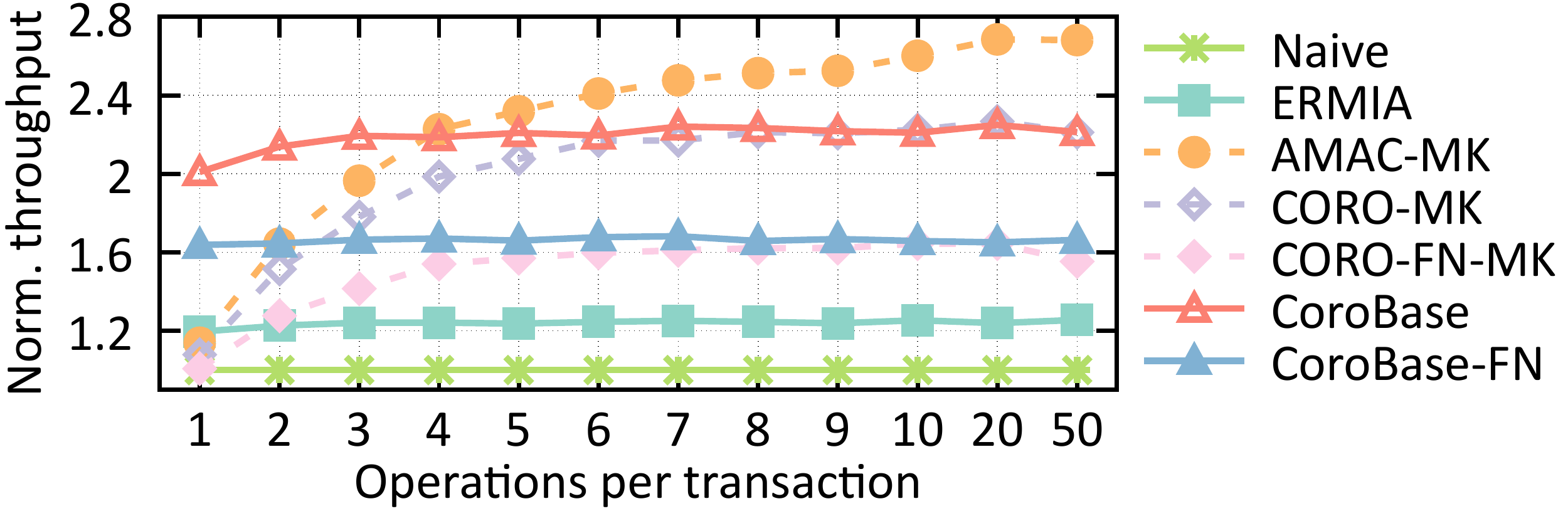}
\caption{\textmd{YCSB read-only performance normalized to \naive under 48 threads.
\cb also benefits from prefetching for very short transactions (1--4 operations) while AMAC and other multi-key based approaches inherently cannot under thread-to-transaction.}}
\label{fig:txn-size-read-only}
\end{figure}

Coroutine-to-transaction also makes it possible for short transactions to benefit from prefetching.
We perform the same YCSB read-only workload but vary the number of reads per transaction. 
As shown in Figure~\ref{fig:txn-size-read-only},  
\cb outperforms all \multiget approaches for very short transactions (1--4 record reads) and continues to outperform all approaches except \amac for larger transactions, as batches are formed across transactions.

These results show that inter-transaction batching enabled by coroutine-to-transaction can match closely the performance of intra-transaction batching while retaining backward compatibility.
Hyperthreading helps to a limited extent and \cb can extract more performance, partially due to the limited hardware contexts (two per core) available in modern Intel processors.
Compared to prior approaches, CoroBase and coroutine-to-transaction further enable short transactions (with little/no intra-transaction batching opportunity) to also benefit from software prefetching. 


\subsection{Write and Scan Workloads}
\label{subsec:eval-write}

\begin{figure}[t]
\centering
\includegraphics[width=\columnwidth]{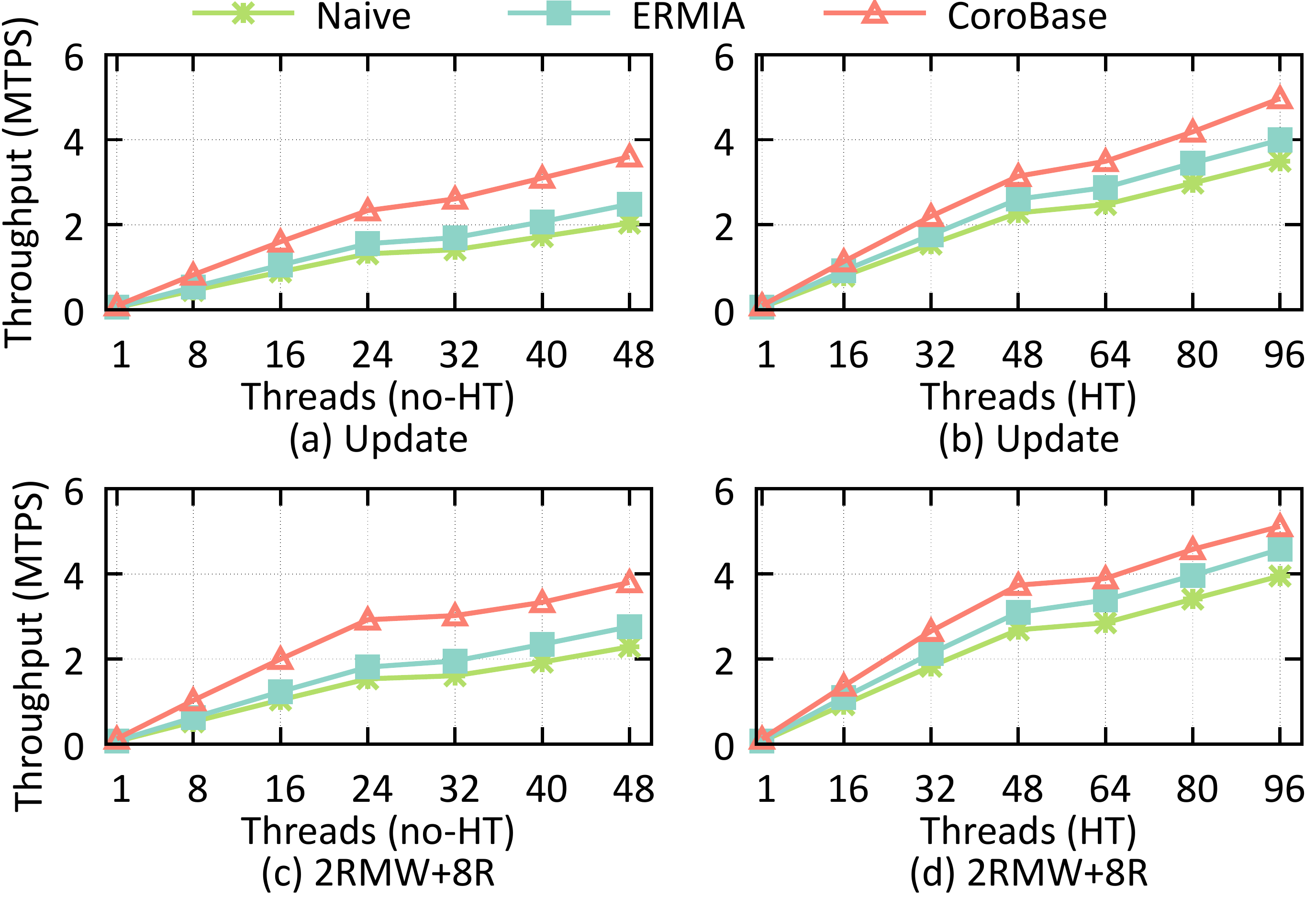}
\caption{\textmd{Throughput of an update-only YCSB workload with 10 blind writes per transaction (a--b), and a mixed workload (c--d).}}
\label{fig:update-rmw}
\end{figure}


One of our goals is to use software prefetching to hide data stalls as much as possible.
While most prior work focused on read-only or read-dominant operations, we extend our experiments to cover write-intensive scenarios.
Write operations also need to traverse index and version chains to reach the target record where data stalls constitute a significant portion of stalled cycles. 
Figure~\ref{fig:update-rmw}(a--b) plots the throughput of update-only (blind write) workload, where each transaction updates 10 uniform randomly-chosen records. 
\cb achieves up to 1.77$\times$ and 1.45$\times$ higher throughput than \naive and \ermia, respectively.
We observe similar results but with lower improvement for a mixed workload that does two RMWs and eight reads per transaction, shown in Figure~\ref{fig:update-rmw}(c--d); a pure RMW workload showed similar trends (not shown here).
The lower improvement comes from the fact that the read operation before each modify-write has already brought necessary data into CPU caches, making subsequent transaction switches for modify-write pure overhead.
Moreover, compared to read operations, write operations need to concurrently update the version chain using atomic instructions, which cannot benefit much from prefetching.
Nevertheless, the results show that even for write-intensive workloads, prefetching has the potential of improving overall performance as data stalls still constitute a significant portion of total cycles.

\begin{figure}[t]
\centering
\includegraphics[width=\columnwidth]{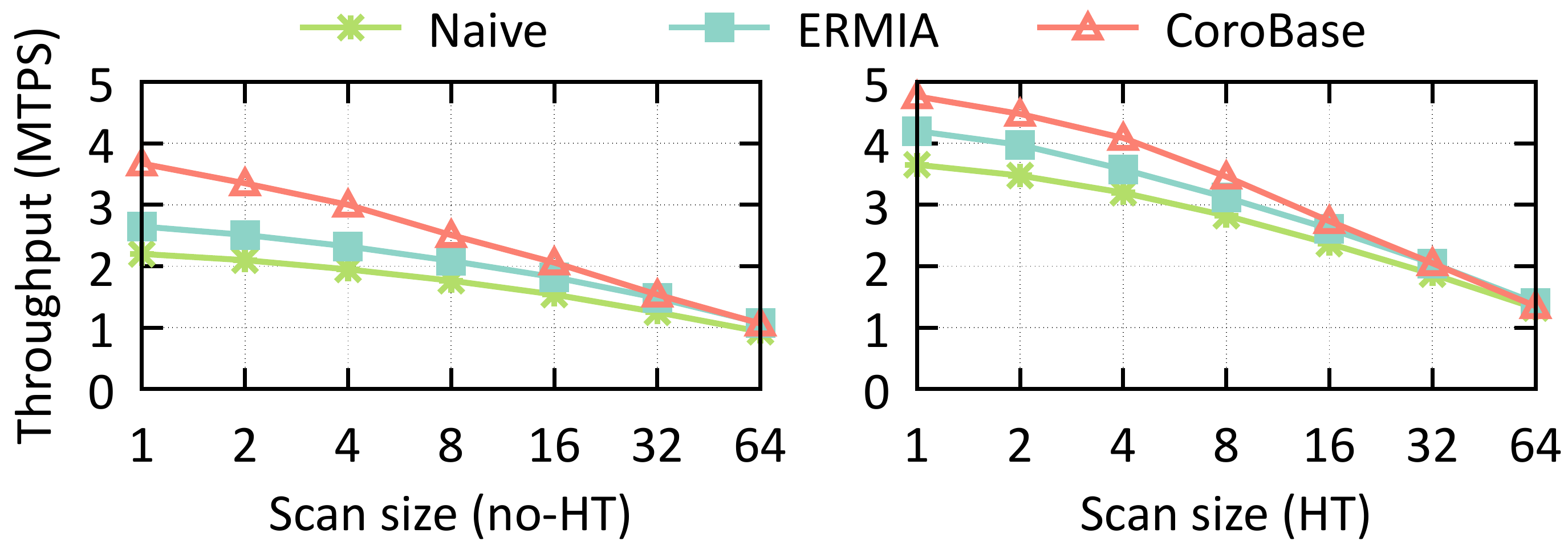}
\caption{\textmd{Throughput of a YCSB scan workload under 48 threads (left) and 96 hyperthreads (right). 
The benefits of prefetching diminish with larger scan sizes: more records can be directly retrieved at the leaf level using B-link-tree structures. }}
\label{fig:scan-only}
\end{figure}

Unlike point-read operations, we observe that scan operations do not always benefit as much. 
Figure \ref{fig:scan-only} shows the throughput of a pure scan workload. 
As we enlarge the number of scanned records, the performance of \naive, \ermia and \cb converges. 
This is because Masstree builds on a structure similar to B-link-trees~\cite{BlinkTree} where border nodes are linked, minimizing the need to traverse internal nodes.
It becomes more possible for the border nodes to be cached and with longer scan ranges, more records can be retrieved directly at the leaf level, amortizing the cost of tree traversal.

\subsection{Impact of Key Lengths and Data Sizes}
\begin{figure}[t]
\centering
\includegraphics[width=\columnwidth]{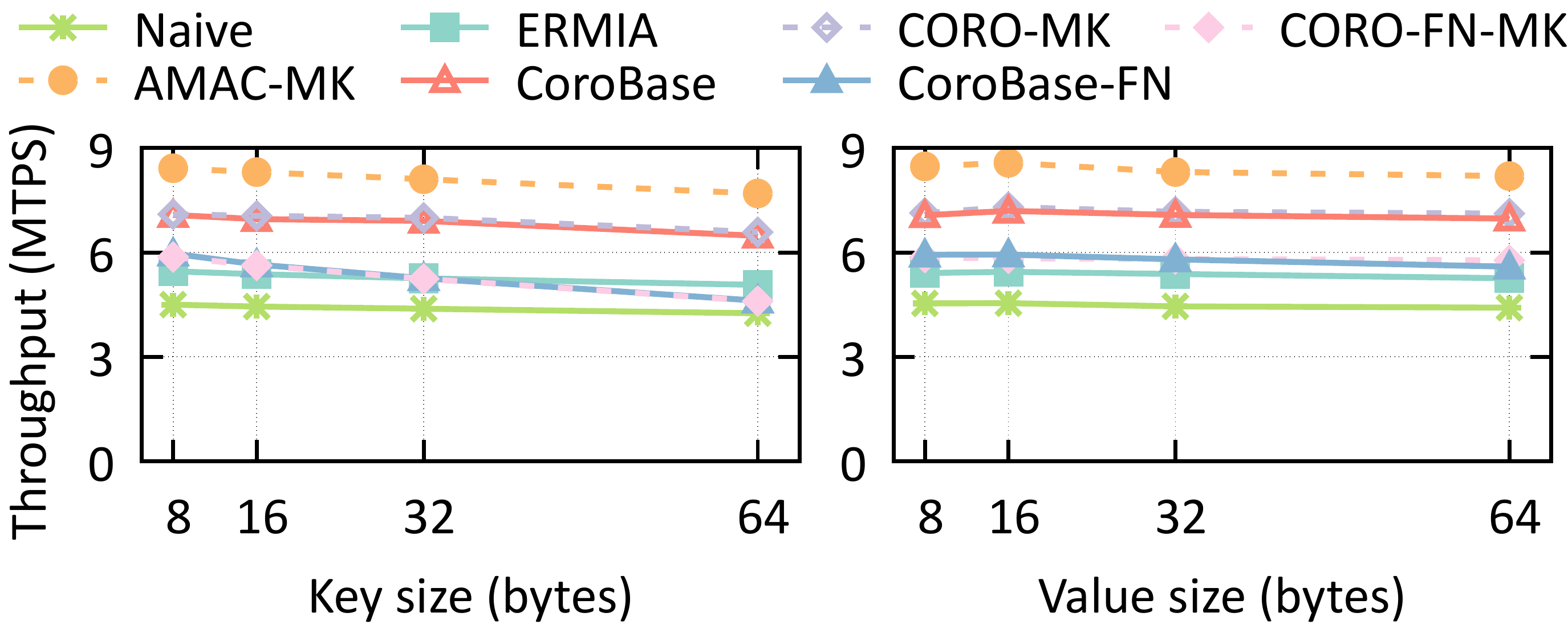}
\caption{\textmd{Throughput of YCSB read-only workload (10 reads per transaction) under varying key/value sizes and 96 hyperthreads.}}
\label{fig:key-value-length}
\end{figure}
\begin{figure}[t]
\centering
\includegraphics[width=\columnwidth]{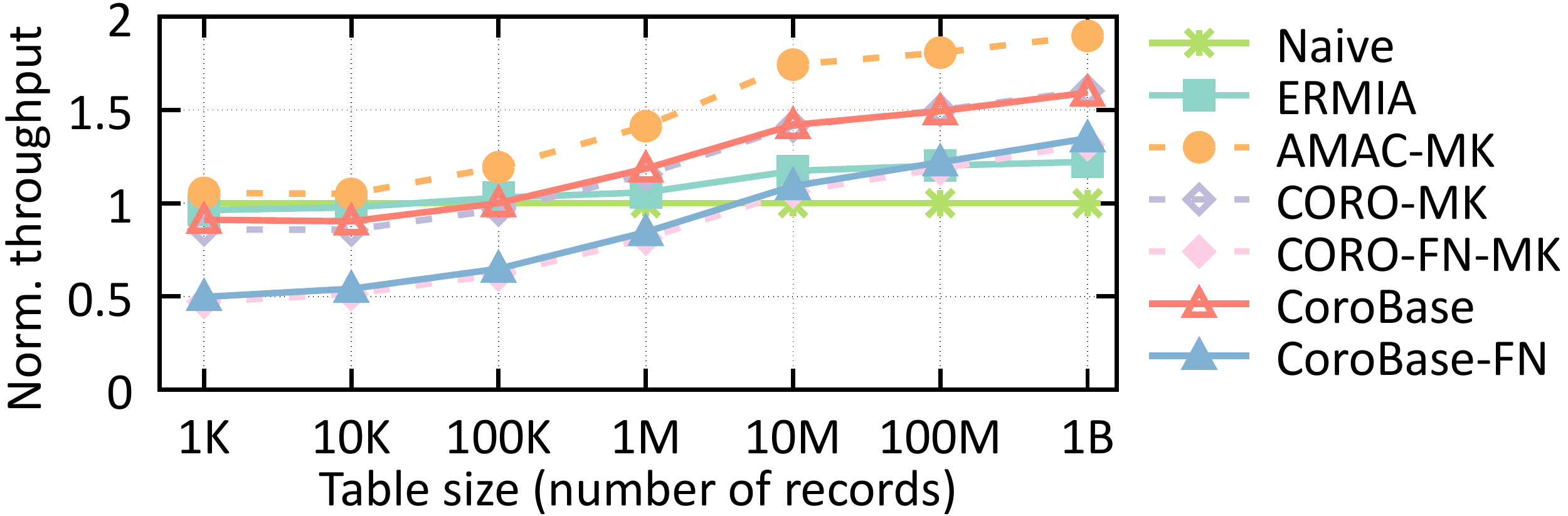}
\caption{\textmd{YCSB read-only (10 reads per transaction) performance normalized to \naive under 96 hyperthreads and varying table sizes.}}
\label{fig:data-size}
\end{figure}

Now we examine how key length, value size, and database size affect runtime performance. 
Figure~\ref{fig:key-value-length}(left) shows the throughput of the YCSB read-only workload under different key lengths.
\cb retains high performance across different key lengths, but \cbfn performs worse under longer keys which cause more coroutine calls to drill down Masstree's hybrid trie/B-tree structure, causing high switching overhead.
As shown in Figure \ref{fig:key-value-length}(right), 
value size does not impact the overall relative merits of different variants. 
Figure \ref{fig:data-size} depicts the impact of data size by plotting the throughput normalized to that of \naive.
As shown, with small table sizes, interleaving does not improve performance.
For example, with 10K records, the total data size (including database records, index, etc.) is merely 1.23MB, whereas the CPU has 35MB of last level cache, making all the anticipated cache misses cache hits.
Context switching becomes pure overhead and exhibits up to $\sim$12.64\% lower throughput (compared to \naive). 
Approaches that use fully-nested coroutines (\cbfn and \corofn) exhibit even up to 50\% slower performance, whereas other approaches including \cb keep a very low overhead. 
In particular, \cb follows the trend of \amac with a fixed amount of overhead.

These results again emphasize the importance of reducing switching overhead.
\cb's low switching overhead for longer keys and small tables make it a practical approach for systems to adopt. 
We also expect coroutine support in compilers and runtime libraries to continue to improve and reduce switching overhead in the future.

\subsection{Impact of Skewed Accesses}
\begin{figure}[t]
\centering
\includegraphics[width=\columnwidth]{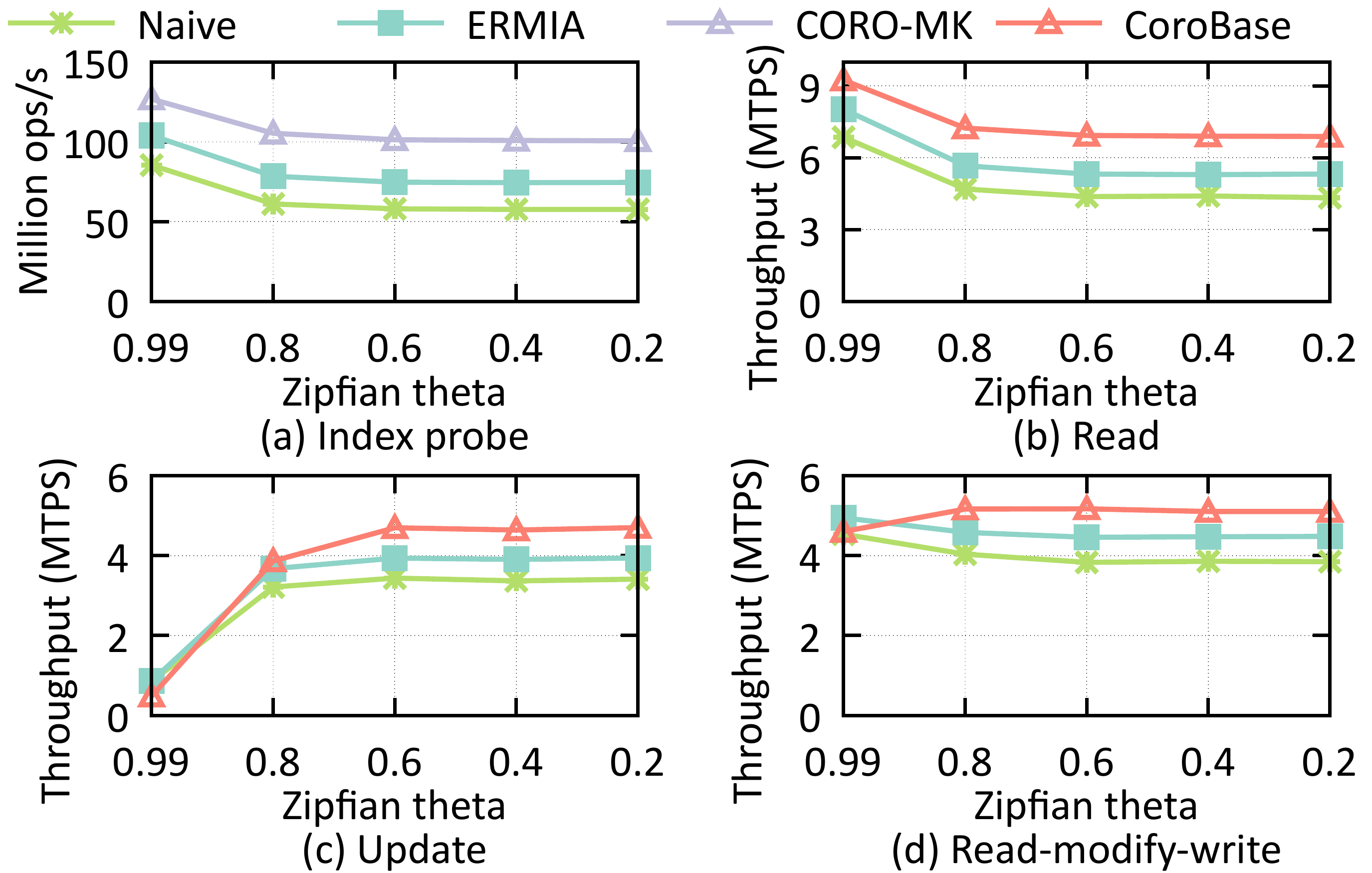}
\caption{\textmd{YCSB throughput under 96 hyperthreads and varying skewness (larger theta indicates more skewed access).}}
\label{fig:zipfian}
\end{figure}
Our last microbenchmark tests how different approaches perform under varying skewness.
Figure~\ref{fig:zipfian}(a) calibrates the expectation using a multi-get based index probing workload that does not access records.
Each transaction here issues 10 operations.
With higher skewness, all schemes perform better because of the better locality.
The YCSB read-only workload in Figure~\ref{fig:zipfian}(b) shows a similar trend.
For update and RMW operations shown in Figures~\ref{fig:zipfian}(c--d), highly skewed workloads lead to high contention and low performance across all schemes, and memory stall is no longer the major bottleneck.
\cb therefore shows lower performance compared to \ermia as switching becomes pure overhead.


\subsection{End-to-End TPC-C Results}
\label{subsec:tpcc}
\begin{figure}[t]
\centering
\includegraphics[width=\columnwidth]{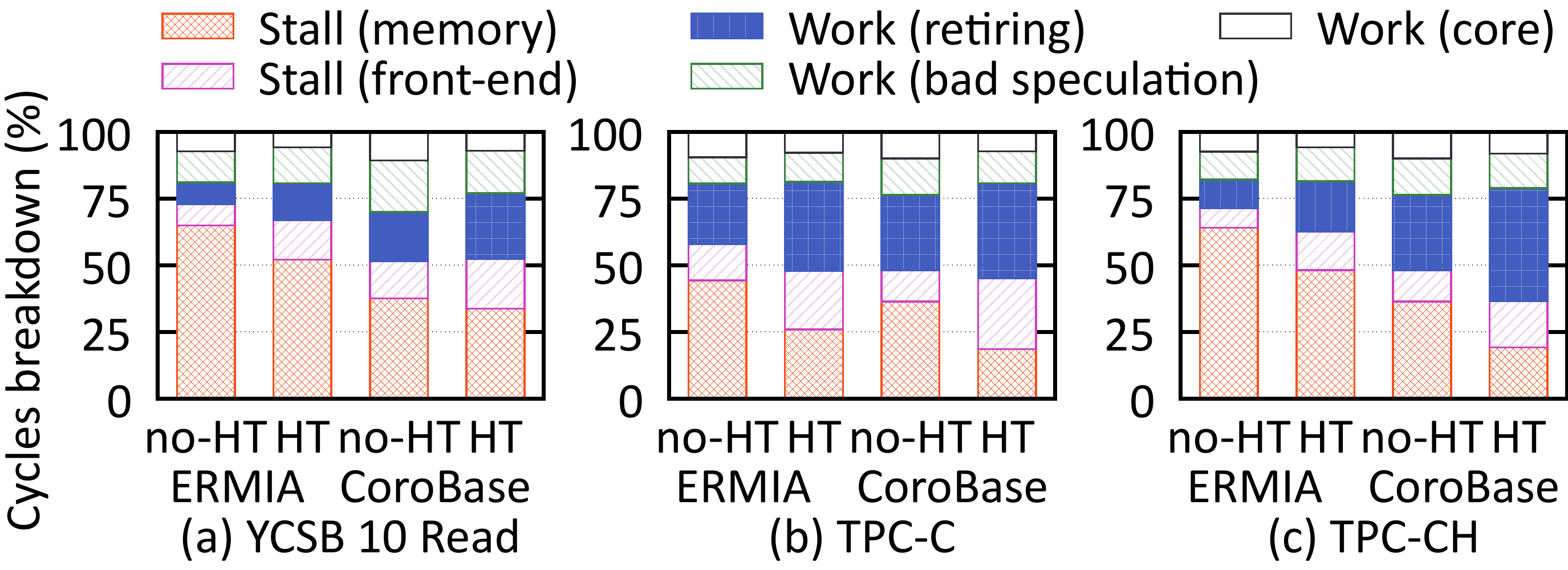}
\caption{\textmd{Microarchitecture analysis for YCSB, TPC-C and TPC-CH workloads. 
\corobase reduces memory stall cycles under all workloads, especially the read-dominant ones.}}
\label{fig:cycles-breakdown}
\end{figure}

\begin{figure}[t]
\centering
\includegraphics[width=\columnwidth]{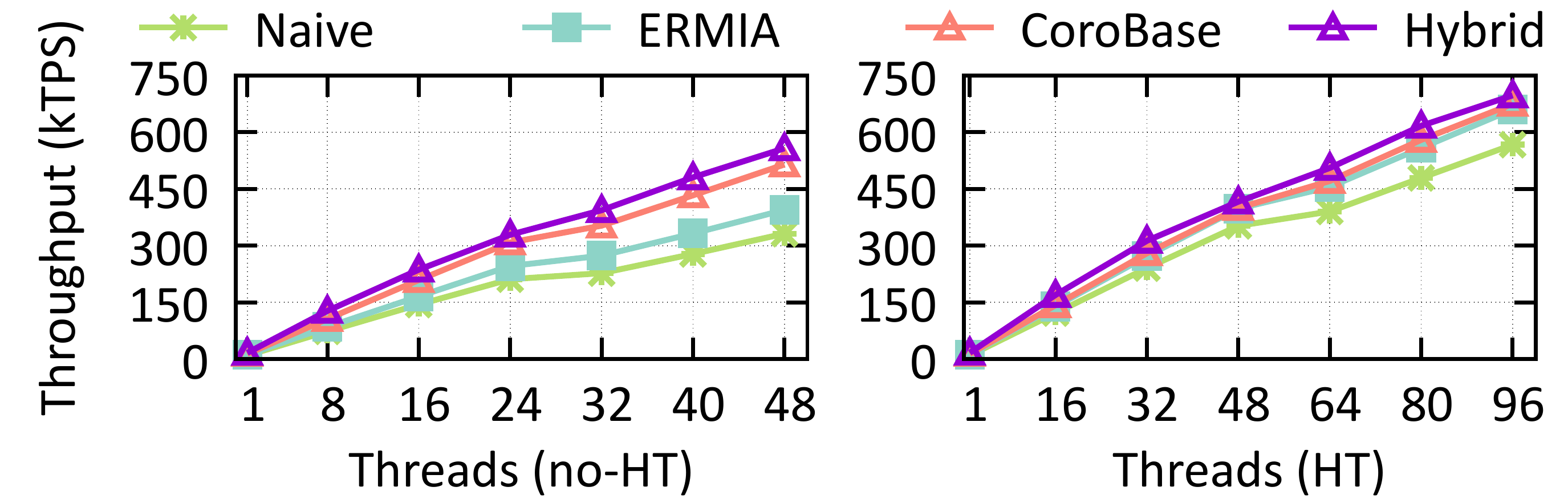}
\caption{\textmd{Throughput of the read-only TPC-CR workload.}}
\label{fig:tpccr}
\end{figure}

Now we turn to TPC-C benchmarks to see how prefetching works in more complex workloads.
We begin with the default TPC-C configuration which is write-intensive.
As shown in Table~\ref{tbl:tpcc}, \corobase manages to perform marginally better than \naive and \ermia without hyperthreading but is 3.4\% slower than \ermia with hyperthreading.
One reason is that TPC-C is write-intensive and exhibits good data locality, with fewer exposed cache misses as shown by the narrow gap between \naive and \ermia.
Our top-down microarchitecture analysis~\cite{IntelManual} result shown in Figure~\ref{fig:cycles-breakdown}(b) verifies this: TPC-C exhibits less than 50\% of memory stall cycles, which as previous work~\cite{InterleaveJoins} pointed out, do not provide enough room to benefit from prefetching.
The high memory stall percentage in Figure~\ref{fig:cycles-breakdown}(a) confirms our YCSB results which showed more improvement.
Under TPC-CR which is read-only, \cb achieves up to 1.55$\times$/1.3$\times$ higher throughput with/without hyperthreading (Figure~\ref{fig:tpccr}).
Notably, with hyperthreading \cb performs similarly (4\% better) to \ermia, showing that using two hyperthreads is enough to hide the memory stalls that were exposed on physical cores for TPC-CR.
Interleaved execution is inherently not beneficial for such workloads, so the goal is for \corobase is to match the performance of sequential execution as close as possible.
\begin{table}[t]
\caption{\textmd{Throughput (TPS) of original TPC-C which is not inherently memory-bound. 
With two-level coroutine, \cb outperforms \naive and matches the performance of \ermia and Hybrid.}}
\label{tbl:tpcc}
\begin{tabular}{@{}lllll@{}}
\toprule
Number of Threads    & \naive   & \ermia & \cb & Hybrid \\ \midrule
48 (no-HT) & 1306197 & 1487147 & 1489567 & 1682290 \\
96 (HT)    & 1985490 & 2195667 & 2120033 & 2197260 \\\bottomrule
\end{tabular}
\end{table}

\begin{figure}[t]
\centering
\includegraphics[width=\columnwidth]{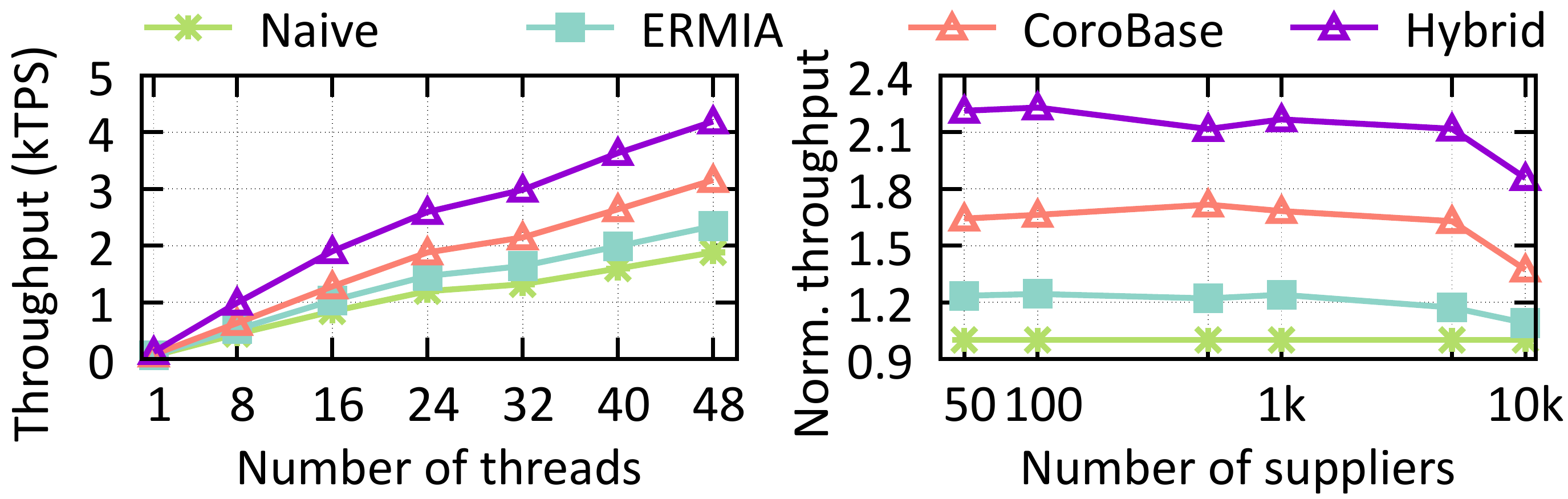}
\caption{\textmd{Scalability (left) and throughput under 48 threads normalized to \naive (right) of TPC-CH.
\corobase reaps more benefits from software prefetching as the workload becomes read-dominant.}}
\label{fig:tpcch}
\end{figure}

Under TPC-CH which is read-intensive and exhibits high memory stall cycles, \cb achieves up to 34\% higher throughput than \ermia when the number of suppliers is set to 100 under 48 threads in Figure~\ref{fig:tpcch}(left).
Figure~\ref{fig:tpcch}(right) explores how throughput changes as the size of the Q2* transaction changes.
\cb exhibits 25--39\% higher throughput than \ermia, and the numbers over \naive are 37--71\%.
Correspondingly, Figure~\ref{fig:cycles-breakdown}(c) shows fewer memory stall cycles under \corobase.

We explore the potential of using selective multi-key operations in \corobase (Section~\ref{subsec:discuss}) with the \hybrid variant. 
We use \multiget coroutines for long queries in NewOrder, StockLevel and Query2 to retrieve items and supplying warehouses, recently sold items and items from certain pairs of the Stock and Supplier tables, respectively.
Other queries use the same single-key operations as in \cb.
As shown in Table~\ref{tbl:tpcc} and Figures~\ref{fig:tpccr}--\ref{fig:tpcch}, \hybrid outperforms \cb by up to 1.29$\times$/1.08$\times$/1.36$\times$ under TPC-C/TPC-CR/TPC-CH, taking advantage of data locality and reduced switching overhead.
The tradeoff is increased code complexity. 
For operators that already exhibit or are easily amenable to multi-key interfaces, \hybrid can be an attractive option.

\subsection{Impact on Transaction Latency}
\begin{figure}[t]
\centering
\includegraphics[width=\columnwidth]{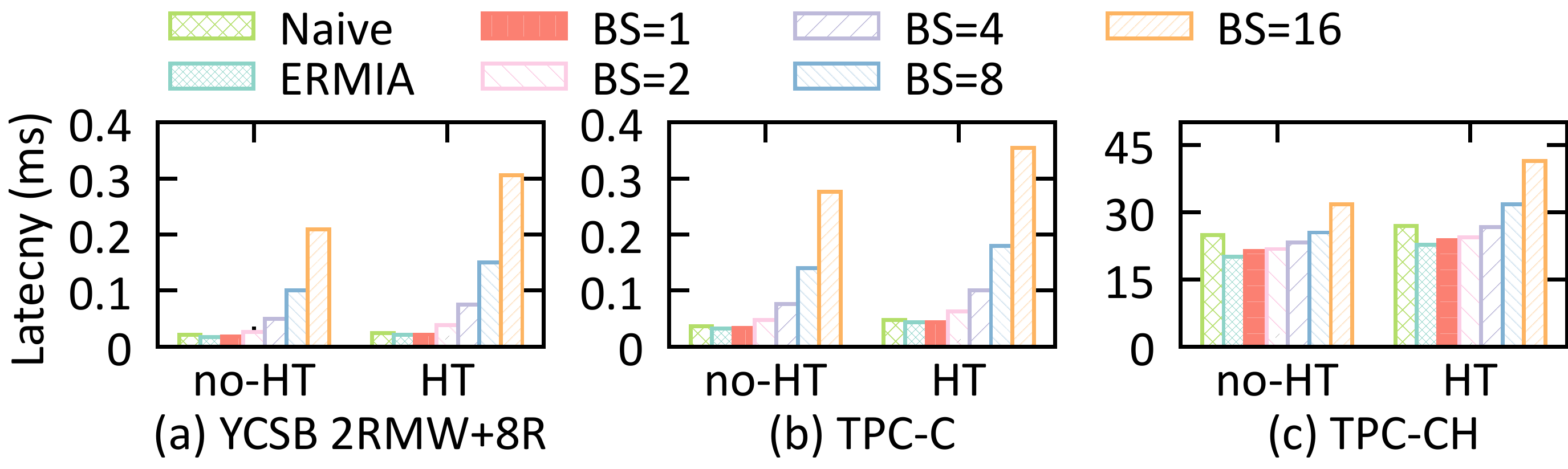}
\caption{\textmd{Average transaction latency under varying batch sizes (BS=1--16) with asynchronous commit. 
The end-to-end impact is expected to be small under group/pipelined commit.
}}
\label{fig:latency-rmw-tpcc-tpcch}
\end{figure}

We analyze the impact of interleaved execution on transaction latency using a mixed YCSB workload (2 RMW +8 read operations per transaction), TPC-C and TPC-CH.
As shown in Figure~\ref{fig:latency-rmw-tpcc-tpcch}, 
with larger batch sizes, transaction latency increases.
We find setting batch size to four to be optimal for the tested YCSB and TPC-C workloads: when batch size exceeds four, latency grows proportionally since there is no room for interleaving. 
TPC-CH exhibits smaller increase in latency, indicating there is much room for overlapping computation and data fetching.
Hyperthreading also only slightly increases average latency.
Note that in this experiment we use asynchronous commit which excludes I/O cost for persisting log records.
Many real systems use group/pipelined commit~\cite{Aether} to hide I/O cost.
The result is higher throughput but longer latency for individual transactions (e.g., 1--5ms reported by recent literature~\cite{QueryFresh}).
Therefore, we expect the increased latency's impact on end-to-end latency in a real system to be very low.

\subsection{Coroutine Frame Management}
Flattening coroutines may increase code size and incur more instruction cache misses (Section~\ref{subsec:2level}).  
For two-level coroutine-to-transaction, the size of a coroutine frame for the flattened read/update/insert functions are 232/200/312 bytes, respectively.
With fully-nested coroutine-to-transaction, five small functions are involved, and their size range from 112--216 bytes, which are indeed smaller than their flattened counterparts.
When handling a request, \corobase allocates a single coroutine frame.
\cbfn maintains a chain of at least five coroutine frames (e.g., for reads it adds up to 808 bytes per request) and switches between them.
Performance drops with both larger memory footprint (therefore higher allocation/deallocation overhead) and switching.
Overall, two-level coroutines ease this problem and achieves better performance.

\section{Related Work}
\label{sec:related-work}
Our work is closely related to prior work on coroutine-based systems, cache-aware optimizations and database engine architectures.

\textbf{Interleaving and Coroutines.}
We have covered most related work in this category~\cite{ChenHashJoin,AMAC,InterleaveJoins,InterleaveJoinsVLDBJ,KillerNanosecond} in Section~\ref{sec:bg}, so we do not repeat here. 
The gap between CPU and the memory subsystem continues to widen with the introduction of more spacious but slower persistent memory~\cite{Intel3DXP}. 
Psaropoulos et al.~\cite{CoroNVM} adapted interleaved execution using coroutine to speed up index joins and tuple construction on persistent memory.
Data stalls are also becoming a bottleneck for vectorized queries~\cite{VecQuery,RethinkSIMDVec}. 
IMV~\cite{InterleaveSIMD} interleaves vectorized code to reduce cache misses in SIMD vectorization.

\textbf{Cache-Aware Optimizations.}
Many proposals try to match in-memory data layout with access pattern for better locality.
CSB+-tree~\cite{CSBTree} stores child nodes of each internal node contiguously to better utilize the cache but trades off update performance.
ART~\cite{ART} is a trie that uses a set of techniques to improve space efficiency and cache utilization.
HOT~\cite{HOT} dynamically adjusts trie node span based on data distributions to achieve a cache-friendly layout. 
Prefetching B+-trees~\cite{pBTree} uses wider nodes to reduce B+-tree height and cache misses during tree traversal.
Fractal prefetching B+-trees~\cite{fpBTree} embed cache-optimized trees within disk-optimized trees to optimize both memory and I/O.
Masstree~\cite{Masstree} is a trie of B+-trees that uses prefetching. 
Software prefetching was also studied in the context of compilers~\cite{CompilerPrefetching}.  
At the hardware level, path prefetching~\cite{PathPrefetching} adds a customized prefetcher to record and read-ahead index nodes. 
Widx~\cite{Widx} is an on-chip accelerator that decouples hashing and list traversal and processes multiple requests in parallel.

\textbf{Database Engine Architectures.}
Most main-memory database engines~\cite{Silo,FOEDUS,ERMIA,Hekaton,Hyper} use the shared-everything architecture that is easy to be adapted to coroutine-to-transaction.
Some systems~\cite{DORA,Bohm,PWV,ReactDB} allow intra-transaction parallelism with delegation. 
Techniques in \corobase are complementary to and can be used to hide data stalls in these systems (Section~\ref{subsec:discuss}).
To adopt coroutine-to-transaction in shared-nothing systems~\cite{HStore}, concurrency control and synchronization need to be re-introduced to allow context switching.
Data stall issues were also identified in column stores~\cite{boncz1999database,CStore,MonetDB} and analytical workloads~\cite{OLAPAnalysis}. 
Exploring ways to hide data stalls in these systems is interesting future work.

\section{Conclusion}
\label{sec:conclusion}
We highlighted the gap between software prefetching and its adoption in database engines. 
Prior approaches often break backward compatibility using multi-key interfaces and/or are piece-wise solutions that optimize individual database operations. 
Leveraging recently standardized lightweight coroutines, we propose a new coroutine-to-transaction execution model to fill this gap. 
Coroutine-to-transaction retains backward compatibility by allowing inter-transaction batching which also enables more potential of software prefetching.
Based on coroutine-to-transaction, we build \corobase, a main-memory database engine that judiciously leverages coroutines to hide memory stalls. 
\corobase achieves high performance via a lightweight two-level coroutine design. 
Evaluation results show that on a 48-core server \corobase is up to $\sim$2$\times$ faster than highly-optimized baselines and remains competitive for workloads that inherently do not benefit from software prefetching.

\noindent\textbf{\\Acknowledgments.} 
We thank Qingqing Zhou and Kangnyeon Kim for their valuable discussions.
We also thank the anonymous reviewers for their useful feedback.
This paper is dedicated to the memory of our wonderful colleague Ryan Shea.

\balance
\bibliographystyle{ACM-Reference-Format}
\bibliography{ref}

\end{document}